\documentstyle[preprint,aps,floats]{revtex}

\begin{document}

\draft

\preprint{\vbox{
\hbox{UASLP-IF-00-05}
\hbox{ESFM/F04/00}
\hbox{hep-ph/00nnmmm}
}}

\title{
Radiative corrections to the semileptonic Dalitz plot with angular
correlation between polarized decaying hyperons and emitted charged
leptons
}
\author{
A.~Mart{\'\i}nez,$^*$ J.~J. Torres,$^*$ Rub\'en
Flores-Mendieta,$^\dagger$ and A.~Garc{\'\i}a$^\ddagger$
}
\address{
$^{*}$Escuela Superior de F{\'\i}sica y Matem\'aticas del IPN, \\ 
Apartado Postal 75-702, M\'exico, Distrito Federal 07738, Mexico \\
$^\dagger$Instituto de F{\'\i}sica, Universidad Aut\'onoma de San Luis
Potos{\'\i}, \\
\'Alvaro Obreg\'on 64, Zona Centro, San Luis Potos{\'\i}, S.L.P. 78000,
Mexico \\
$^{\ddagger}$Departamento de F{\'\i}sica, Centro de Investigaci\'on y de
Estudios Avanzados del IPN, \\
Apartado Postal 14-740, M\'exico, Distrito Federal 07000, Mexico
}

\date{June 23, 2000}

\maketitle

{\tighten
\begin{abstract}
We obtain a model-independent expression for the Dalitz plot of
semileptonic decays of polarized hyperons including radiative corrections
to order $\alpha$ and neglecting terms of order $\alpha q/\pi M_1$, where
$q$ is the four-momentum transfer and $M_1$ is the mass of the decaying
hyperon. We specialize our results to exhibit the correlation between the
charged-lepton momentum and the spin of the decaying hyperon. We present
results for the three-body region of the Dalitz plot and for the complete
Dalitz plot (which includes the four-body region). From these results we
also obtain the corresponding radiative corrections to the integrated
lepton spin-asymmetry coefficient. Our formulas are valid for charged as
well as for neutral decaying hyperons and are appropriate for
model-independent experimental analysis whether the real photon is
discriminated or not.
\end{abstract}
}


\newpage

\tightenlines

\section{Introduction}

The form factors of hyperon semileptonic decays, $A\rightarrow B \ell
\nu_\ell$, contain important information about the low-energy strong
interactions of spin-$1/2$ baryons ($A$ and $B$ are such baryons and
$\ell$ and $\nu_\ell$ are the accompanying charged lepton and neutrino).
Their experimental determination requires the use of accurate formulas in
the analysis of the measurements of several observables. An important one
of these observables is the angular correlation between the spin ${\bf
\hat s}_1$ of $A$ and the direction ${\bf \hat l}$ of the momentum of
$\ell$. It is the purpose of this paper to calculate the radiative
corrections (RC) to the Dalitz plot (DP) with the spin correlation ${\bf
\hat s}_1 \cdot {\bf \hat l}$ exhibited explicitly.

We shall obtain expressions that are suitable for a model-independent
experimental analysis. The model dependence of the virtual RC is handled
following the approach of Sirlin\cite{Sirlin} to the RC of neutron beta
decay whereas the model dependence of the bremsstrahlung is controlled by
the theorem of Low\cite{Low}. In previous work we have discussed the
unpolarized DP~\cite{Tun} and the DP with the ${\bf \hat s}_1 \cdot {\bf
\hat p}_2$ spin correlation kept explicitly~\cite{Flores} (${\bf \hat
p}_2$ is the direction of the momentum of the emitted baryon $B$). It is
not possible to derive the result for the spin correlation ${\bf \hat s}_1
\cdot {\bf \hat l}$ from the final result of Ref.~\cite{Flores}, because
all kinematical integrations, except for the $\ell$ and $B$ energies $E$
and $E_2$, respectively, were already performed. However, since we are
going to follow the same approach of this reference, much of the work has
already been advanced.

The bremsstrahlung RC is a four-body decay whose DP covers entirely the DP
of the three-body decay $A \rightarrow B \ell \nu_\ell$. We shall refer to
the latter as the three-body region (TBR) and to the non-overlap of the
former and the latter as the four-body region (FBR). Even when no
experimental arrangement has been made to detect and discriminate real
photons, it is possible to eliminate the photons that belong to the FBR by
energy-momentum conservation. Therefore, in calculating the bremsstrahlung
RC we shall keep a clear distinction between these two regions.

We shall also obtain the radiative corrections to the integrated lepton
spin-asymmetry coefficient $\alpha_\ell$. As we shall see the distinction
between the TBR and the FBR leads to a perceptible change in the RC to
this asymmetry coefficient.

Our results will be presented in two final forms. One where the triple
integration over the real photon three-momentum $\bf k$ is left indicated
and ready to be performed numerically. And another one, an analytical
form, where such a triple integration has been performed. Both forms can
be used to numerically cross-check on one another. However, the analytical
result, although tedious to feed into a Monte Carlo program, leads to a
considerable saving of computer time, because the triple integration does
not have to be performed within the Monte Carlo every time $E$ and $E_2$
or the form factors are changed.

For the use of our results it is important that this paper be as
self-contained as possible. In Sec. II we introduce our notation and
conventions and we review the virtual RC; also the infrared divergence of
this part is clearly separated. In Sec. III the real photon emission is
calculated and separated into the contributions from the TBR and FBR. The
calculation of Ref.~\cite{Flores} is adapted to the present case. The
infrared divergence is extracted following Ref.~\cite{Ginsberg} and its
cancellation with the one of Sec.~II is discussed in detail. Our first
main result is established, allowing for the elimination (or not) of real
photons from the experimental analysis. In Sec. IV we proceed to the
analytical evaluation of the triple integration over the bremsstrahlung
three-momentum. Our second main result is established, also allowing for
the experimental discrimination (or not) of real photons. In Sec.~V we use
the analytical result to obtain the RC to the asymmetry coefficient
$\alpha_\ell$. In Sec.~VI we make numerical evaluations for several HSD
and also for the $\Lambda_c^+ \rightarrow \Lambda e^+ \nu$ decay. We
compare with other results available in the literature. Section VII is
reserved to discuss and summarize our results. To make this paper
self-contained we introduce three appendices. In Appendix A we give the
amplitudes for virtual and bremsstrahlung RC, emphasizing how the model
dependence is kept under control. In Appendix B we give the analytical
expressions of all the coefficients, both new and of Refs.~\cite{Tun} and
\cite{Flores} required to compute the RC to the ${\bf \hat s}_1 \cdot {\bf
\hat l}$ correlation. Finally, in Appendix C we review briefly the
procedure of Ref.~\cite{Ginsberg} to extract the infrared divergences.

Our results have been obtained neglecting contributions of order $\alpha
q/\pi M_{1}$ and higher ($q$ is the momentum transfer and $M_{1}$ is the
mass of $A$). They cover both neutral and charged $A$ and are reliable up
to $0.5\%$ or better in HSD. Furthermore, they provide a useful result for
charm decay experiments with several thousands of events. For higher
statistics experiments it will be necessary to incorporate $\alpha q/ \pi
M_{1}$ contributions.

\section{\bf Virtual radiative corrections}

In this section we shall discuss the virtual radiative corrections, up to
order $\alpha $ and neglecting terms of order $\alpha q/\pi M_{1}$, to the
DP of the hyperon semileptonic decay
\begin{equation}
A \rightarrow B+ \ell + \nu_\ell , \label{eq:e1}
\end{equation}
with $A$ polarized. Our results will be especialized to exhibit the
angular correlation ${\bf \hat s}_1 \cdot {\bf \hat l}$.

A convenient procedure to get such results is that of Ref.~\cite{Flores}.
So, in this paper we adopt the same approach, the same approximations and
the same conventions of this reference. In this way, $p_1 = \left( E_1,
{\bf p}_1 \right)$, $p_2 = \left( E_2, {\bf p}_2 \right)$, $l = \left( E,
{\bf l} \right)$, and $p_{\nu}=\left( E_{\nu}, {\bf p}_\nu \right)$ will
be the four-momenta of $A$, $B$, $\ell$, and $\nu_\ell$, respectively.
$M_1$, $M_2$, and $m$ will denote the masses of the first three particles.
We shall assume throughout this paper that $m_\nu=0$. ${\bf \hat p}_2$
will denote a unit vector along ${\bf p}_2$, etc. We shall make our
calculations in the center-of-mass frame of $A$. In this case, $p_2$, $l$,
and $p_\nu$ will denote the magnitudes of the corresponding three-momenta.
There will be no confusion because the expressions obtained will not be
manifestly covariant. Because we want our results to exhibit the
correlation ${\bf \hat s}_1 \cdot {\bf \hat l}$, it is convenient to
choose the $z$-axis along ${\bf l}$ and not along ${\bf p}_2$ as in
Ref.~\cite{Flores}.

At this point, it is convenient to mention that it is not possible to
obtain the virtual RC of our present case by using the final virtual RC
given by $d\Gamma_V$ of Eq.~(15) of Ref.~\cite{Flores}. This is because in
that equation the correlation ${\bf \hat s}_1 \cdot {\bf \hat p}_2$ was
singled out after the integration over the azimuthal angle $\phi_\ell$ of
${\bf l}$ was performed [in such Eq.~(15) $d\phi_\ell$ is still present in
the phase space. As it appears and because of the choice of $z$-axis its
integration amounts a $2\pi$ factor]. Therefore, all the terms in
$d\Gamma_V$ containing the product ${\bf \hat s}_1 \cdot {\bf \hat l}$,
and which appear before the integration over $d\phi_\ell$ is performed,
have been transformed leaving the correlation ${\bf \hat s}_1 \cdot {\bf
\hat p}_2$ only. There is no way to recover the ${\bf \hat s}_1 \cdot {\bf
\hat l}$ terms from this Eq.~(15) of Ref.~\cite{Flores}. So, for obtaining
our $d\Gamma_V$, exhibiting the correlation ${\bf \hat s}_1 \cdot {\bf
\hat l}$ only, we have to take a few steps back before that Eq.~(15).

Our calculation now starts at the point where the scalar products ${\bf
\hat s}_1 \cdot {\bf \hat l}$ and ${\bf \hat s}_1 \cdot {\bf \hat p}_2$
appear for the first time, that is, at Eq.~(13) of Ref.~\cite{Flores},
namely,
\begin{equation}
\sum_{{\rm spins}} \left| M_V \right|^2 = \frac{1}{2}\sum_{{\rm spins}}
\left| M_V^\prime \right|^2-\frac{1}{2}\sum_{{\rm spins}}\left|
M_V^{(s)} \right|^2 . \label{eq:e2}
\end{equation}

Here $M_V$ is the sum of the order zero amplitude $M_0^\prime$,
corrected by the model-dependent part of the virtual radiative corrections
through the modified form factors $f_1^\prime$ and $g_1^\prime$, and
the model-independent amplitude $M_v$ of such RC. $M_0^\prime$ and
$M_v$ are given explicitly in Appendix A [see Eqs.~(\ref{eq:eA1}) and
(\ref{eq:eA4})].

The two terms on the right-hand side of Eq.~(\ref{eq:e2}) arise after the
spinor $u_{A}\left( p_{1}\right) $ of the polarized hyperon $A$ is
replaced by $\Sigma (s_1) u_A \left( p_{1}\right)$ in the amplitude $M_V$.
$\Sigma (s_1)$ is the spin projection operator of $A$ given in Eq.~(4) of
Ref.~\cite{Flores}.

With Eq.~(\ref{eq:e2}) we can express the differential decay rate
$d\Gamma_V$ as
\begin{eqnarray}
d\Gamma_V &=&\frac{dE_2 \,\,dE\,\, d\Omega_\ell \,\, d\phi_2}{\left( 2\pi
\right)^{5}}M_2 m m_\nu \left[ \frac{1}{2}\sum_{{\rm spins}}\left|
M_V^\prime \right|^2-\frac{1}{2}\sum_{{\rm spins}}\left|
M_V^{(s)} \right|^2 \right] \nonumber \\ 
& = & d\Gamma_V^\prime-d\Gamma_V^{(s)} . \label{eq:e3}
\end{eqnarray}
Notice the variables in the phase space of this equation, they correspond
to our new choice of the $z$-axis along ${\bf l}$.

In Eq.~(\ref{eq:e3}), $d\Gamma_V^\prime$ corresponds to the first
term within the square brackets. It can be identified with the
differential decay rate with virtual radiative corrections of unpolarized
$A$ given in Eq.~(10) of Ref.~\cite{Tun} and, therefore, there is no need
to recalculate it now. $d\Gamma_V^{(s)}$ corresponds to the second term
of the square brackets of Eq.~(\ref{eq:e3}) and it contains the scalar
products ${\bf \hat s}_1 \cdot {\bf \hat l}$, ${\bf \hat s}_1 \cdot {\bf
\hat p}_\nu$, and ${\bf \hat s}_1 \cdot {\bf \hat p}_2$. The scalar
product ${\bf \hat s}_1 \cdot {\bf \hat p}_\nu$ can be expressed in terms
of ${\bf \hat s}_1 \cdot {\bf \hat l}$ and ${\bf \hat s}_1 \cdot {\bf
\hat p}_2$ by three-momentum conservation. In this way, only ${\bf \hat
s}_1 \cdot {\bf \hat l}$ and ${\bf \hat s}_1 \cdot {\bf \hat p}_2$ will
appear in $d\Gamma_V^{(s)}$. Now, we require that ${\bf \hat s}_1 \cdot
{\bf \hat l}$ be the only scalar product present in $d\Gamma_V^{(s)}$.
This can be accomplished by noting that the most general form of ${\bf
\hat s}_1 \cdot {\bf \hat p}_2$ depends on ${\bf \hat s}_1 \cdot {\bf \hat
l}$, ${\bf \hat l}\cdot {\bf \hat p}_2$, and $\cos \phi_2$. The terms
directly proportional to this cosine drop out after integration over
$\phi_2$ from $0$ to $2\pi$. This fact allows us to use the replacement
\begin{equation} 
{\bf \hat s}_1 \cdot {\bf \hat p}_2 \rightarrow ({\bf \hat s}_1 \cdot
{\bf \hat l}) ({\bf \hat l} \cdot {\bf \hat p}_2) = {\bf \hat s}_1 \cdot
{\bf \hat l} \, y_0 , \label{eq:e4}
\end{equation} 
in $d\Gamma_V^{(s)}$, leaving us with an expression which only contains the
correlation ${\bf \hat s}_1 \cdot {\bf \hat l}$. In Eq.~(\ref {eq:e4})
$y_{0}$ is the cosine of the angle between ${\bf \hat p}_2$ and ${\bf \hat
l}$ when the real photon is not present. In the CM of $A$, $y_0$ is given
by
\begin{equation} y_0 = \frac{({E_\nu^0})^2 - p_2^2 - l^2}{2 p_2 l} ,
\label{eq:e5}
\end{equation}
where 
\begin{equation} E_\nu^0 = M_1 - E_2 - E , \label{eq:e6}
\end{equation} is
the neutrino energy (in this three-body decay).

With these considerations in mind, we can express the $d\Gamma_V$ of
Eq.~(\ref{eq:e3}) as
\begin{eqnarray}
d\Gamma_V & = & \frac{G_V^2}{2} \frac{dE_2 \,\, dE \,\, d\Omega_\ell
\,\, d\phi_2}{\left( 2\pi \right)^5} 2M_1 \left\{A_0^\prime +
\frac{\alpha}{\pi}\left( A_{1}^\prime \phi +A_{1}^{\prime
\prime} \phi^\prime \right) \right.  \nonumber \\ & & \mbox{} - \left.
{\bf \hat s}_1 \cdot {\bf \hat l}\left[ B_{0}^{\prime \prime} +
\frac{\alpha}{\pi}\left( B_{2}^\prime \phi + B_{2}^{\prime \prime}
\phi^\prime \right) \right] \right\} \, . \label{eq:e7}
\end{eqnarray}

This $d\Gamma_V$ is the DP with virtual radiative corrections up to order
$\alpha$ (and neglecting terms of order $\alpha q/\pi M_1$), leaving $E_2$
and $E$ as the relevant variables and with only the angular correlation
${\bf \hat s}_1 \cdot {\bf \hat l}$ explicitly exhibited. The integration
over $\phi_2$ only amounts the factor $2\pi$. Now, $A_0^\prime$,
$A_1^\prime$, $A_{1}^{\prime \prime}$ are given in
Eqs.~(\ref{eq:eB1})-(\ref{eq:eB3}) of Appendix B. The new terms
$B_{0}^{\prime \prime}$, $B_{2}^\prime$, $B_{2}^{\prime \prime}$ are
\begin{eqnarray}
B_0^{\prime \prime} & = & Q_6 E p_2 y_0 + Q_7 E l , \label{eq:e8} \\
B_2^\prime & = & -D_3 E_\nu^0 l + D_4 E \left( p_2 y_0 + l \right) ,
\label{eq:e9} \\
B_2^{\prime \prime} & = & D_4 E\left( p_2 y_0 + l \right) .
\label{eq:e10}
\end{eqnarray}

In these equations the coefficients $Q_6$ and $Q_7$ are long quadratic
functions of the form factors. They are given in Eqs.~(B6) and (B7),
respectively, of Ref.~\cite{Flores}. For completeness we repeat them too
[see Eqs.~(\ref{eq:eB9}) and (\ref{eq:eB10})]. $D_3$ and $D_4$ depend on
the leading form factors $f_1^\prime$ and $g_1^\prime$ and they are
given in Eqs.~(\ref{eq:eB13}) and (\ref{eq:eB14}), respectively. The
primes on $A_0^\prime$, $A_1^\prime$, $A_1^{\prime \prime}$,
$B_0^{\prime \prime}$, $B_2^\prime$, and $B_2^{\prime \prime}$
indicate that these terms contain the model-dependence of the virtual
radiative corrections through the leading form factors. The
model-independent functions $\phi $ and $\phi^\prime$ are
\begin{eqnarray}
\phi \left( E\right) &=& 2\left( \frac{1}{\beta}\text{arctanh}\beta
-1\right) \ln \frac{\lambda}{m}-\frac{1}{\beta}\left( \text{arctanh}\beta
\right)^{2}+\frac{1}{\beta}L\left( \frac{2\beta}{1+\beta}\right) 
\nonumber \\
&& \mbox{} + \frac{1}{\beta}\text{arctanh}\beta -\frac{11}{8}+\left\{ 
\begin{array}{lll}
\pi^2/\beta +\frac{3}{2}\ln \left( M_{2}/m\right) &  & \text{NDH} 
\label{eq:e11} \\ 
\frac{3}{2}\ln (M_1/m) &  & \text{CDH}
\end{array}
\right.
\end{eqnarray}
\begin{equation}
\phi^\prime \left( E\right) =\left( \beta -\frac{1}{\beta}\right) 
\text{arctanh}\beta . \label{eq:e12}
\end{equation}

NDH corresponds to neutral decaying hyperons and CDH corresponds to
charged decaying hyperons. The infrared divergence appears in the function
$\phi \left( E\right)$ through the parameter $\lambda$. Let us mention
that $d\Gamma_V$ of Eq.~(\ref{eq:e7}) contains two infrared divergent
terms. The first one appears in $A_1^\prime \phi$ of the
spin-independent part and the other one appears in $B_{2}^\prime \phi$
of the spin-dependent part. Both terms will be canceled by their
counterparts in the bremsstrahlung contribution.

Let us close this section by comparing the $d\Gamma_V$ of
Eq.~(\ref{eq:e7}) with the $d\Gamma_V$ of Eq.~(15) of Ref.~\cite{Flores}.
In spite of minor differences in their phase space factors, we can see
that their spin-independent parts are the same. This is not the case for
their spin-dependent parts. We can appreciate that the coefficients
$A_0^{\prime \prime}$, $A_2^\prime$, and $A_2^{\prime \prime}$,
which appear in the spin-dependent part of $d\Gamma_V$ of
Ref.~\cite{Flores}, have changed to the coefficients $B_0^{\prime 
\prime}$, $B_2^\prime$, and $B_2^{\prime \prime}$, respectively, of
$d\Gamma_V$ of Eq.~(\ref{eq:e7}). We observe in Eqs.~(\ref{eq:e8})-(\ref
{eq:e10}) that in the $B$ coefficients $y_{0}$ always appears as a factor
of $p_{2}$, while for the $A$ coefficients of Ref.~\cite{Flores} $y_{0}$
always appears as a factor of $l$. This latter observation may induce us
to think into the possibility of obtaining the $d\Gamma_V$ of
Eq.~(\ref{eq:e7}) from the $d\Gamma_V$ of Ref.~\cite{Flores} by simply
interchanging $p_2$ with $l$. Unfortunately this rule does not work
because under such an interchange the $A$ coefficients do not lead to the
$B$ coefficients and, thus, we can not obtain Eq.~(\ref{eq:e7}) directly
from the final $d\Gamma_V$ of Ref.~\cite{Flores}.

\section{Bremsstrahlung radiative corrections}

In addition to the virtual RC the bremsstrahlung contributions must be
calculated to get the complete RC to the DP of polarized decaying
hyperons. In this section, we shall obtain them, in the same order of
approximation as the virtual RC, both for the TBR and for the FBR. First,
we shall define those regions and next, we shall proceed to the
calculations. As is discussed in Appendix A, these corrections are
model-independent by virtue of the theorem of Low~\cite{Low}.

\subsection{Kinematics, TBR, and FBR}

The DP in the variables $E$ and $E_{2}$ is the kinematically allowed
region of the four-body decay
\begin{equation}
A \rightarrow B+ \ell +\nu_\ell + \gamma , \label{eq:e13}
\end{equation}
where $\gamma$ represents a real photon with four-momentum
$k=\left(\omega, {\bf k}\right)$. The DP can be seen as the union of the
TBR and FBR, each one defined presently.

The TBR of the DP is the region where the three-body decay~(\ref{eq:e1})
and the four-body decay~(\ref{eq:e13})  overlap completely. The energies
$E$ and $E_{2}$ satisfy the bounds
\begin{equation}
E_{2}^{-}\leq E_{2}\leq E_{2}^{+} , \label{eq:e14}
\end{equation}
and
\begin{equation}
m\leq E\leq E_{m} . \label{eq:e15}
\end{equation}
Here, $E_2^+$ $(E_2^-)$ is the upper (lower) boundary of the TBR given by
\begin{equation}
E_{2}^{\pm}=\frac{1}{2}\left( M_{1}-E\pm l \right) + 
\frac{M_{2}^{2}}{2\left( M_{1}-E\pm l \right)} , \label{eq:e16}
\end{equation}
where $E_m$ is the maximum energy of the charged lepton 
\begin{equation}
E_{m}=\frac{M_{1}^{2}-M_{2}^{2}+m^{2}}{2M_{1}} . \label{eq:e17}
\end{equation}

The FBR of DP is the region where only the four-body decay~(\ref{eq:e13}) 
can take place. The energies $E$ and $E_{2}$ in this region satisfy the
bounds
\begin{equation}
M_2 \leq E_{2}\leq E_{2}^{-} , \label{eq:e18}
\end{equation}
\begin{equation}
m \leq E\leq E_{B} , \label{eq:e19}
\end{equation}
where 
\begin{equation}
E_B = \frac{\left(M_1-M_2\right)^2+m^2}{2\left(M_1-M_2\right)} .  
\label{eq:e20}
\end{equation}
Both regions TBR and FBR are depicted in Fig.~1 of Ref.~\cite{Tun} where
more details about these regions can be found.

We can now proceed to the calculation of the bremsstrahlung RC to the DP.
First, we shall do this for the TBR and next for the FBR. The complete
bremsstrahlung RC to the DP are obtained by simply adding the results of
the TBR to those of the FBR. To obtain the complete RC of each region, we
must also add the virtual RC of Eq.~(\ref{eq:e7}).

\subsection{TBR bremsstrahlung RC}

Here we shall obtain the bremsstrahlung RC restricted to the TBR and with
the angular correlation ${\bf \hat s}_1 \cdot {\bf \hat l}$ explicitly
shown. We shall follow the procedure introduced in Ref.~\cite{Flores}.
Therefore, we start from Eq.~(31) of this reference,
\begin{equation}
\sum_{{\rm spins}}\left| M_{B}\right|^{2}= \frac{1}{2}\sum_{{\rm
spins}}\left| M_{B}^\prime \right|^{2}-\frac{1}{2}\sum_{{\rm
spins}}\left| M_{B}^{(s)}\right|^{2} . \label{eq:e21}
\end{equation}

This equation is the square, summed over spins, of the bremsstrahlung
transition amplitude $M_{B}$ of the four-body process (13) with the spinor
$u_A (p_1)$ replaced by $\Sigma (s_1) u_A (p_1)$. The explicit form of
$M_B$ is given in Eq.~(\ref{eq:eA6}) of Appendix~A.
Equation~(\ref{eq:e21}) enables us to express the bremsstrahlung
differential decay rate as
\begin{eqnarray}
d \Gamma_B^{\rm TBR} & = & \frac{M_{2} m\,m_{\nu}}{\left( 2\pi
\right)^{8}} \frac{d^{3}p_{2}}{E_{2}}\frac{d^3 l}{E}\frac{d^{3}k}{2\omega}
\frac{d^{3}p_{\nu}}{E_{\nu}}\sum_{{\rm spins}} \left|
M_{B}\right|^{2}\delta^{4}\left( p_1 -p_2 - l - p_\nu - k \right)  
\nonumber \\
&\equiv & d\Gamma_B^\prime-d\Gamma_B^{(s)} , \label{eq:e22}
\end{eqnarray}
where $d\Gamma_B^\prime$ contains the first term of Eq.~(\ref{eq:e21}) and
is independent of ${\bf \hat s}_1$, while $d\Gamma_B^{(s)}$ contains the
second term of this equation and is spin-dependent. $d\Gamma_B^\prime$
is readily identified with the bremsstrahlung differential decay rate for
unpolarized hyperons given in Eq.~(33) of Ref.~\cite{Flores} as
\begin{equation}
d\Gamma_B^\prime = d\Gamma_B^{\rm ir}+d\Gamma_B^a + d\Gamma_B^b , 
\label{eq:e23}
\end{equation}
where, in accordance to Eq.~(34) of this reference, 
\begin{equation}
d\Gamma_B^{\rm ir} = \frac{\alpha}{\pi} d\Omega^\prime \left\{
A_{1}^\prime \left[ {\hat I}_{0}\left( k/\lambda \right)
+C+C_{1}\right] +C_{2}\right\} . \label{eq:e24}
\end{equation}

The phase space factor in Eq.~(\ref{eq:e24}) is now 
\begin{equation}
\frac{\alpha}{\pi}d\Omega^\prime = \frac{\alpha}{\pi} \frac{G_V^2}{2}
\frac{dE_2 \, dE \, d\Omega_\ell \, d\phi_2}{\left( 2\pi \right)^{5}}
2M_{1} , \label{eq:e25}
\end{equation}
instead of the $d\Omega$ of Eq.~(35) of Ref.~\cite{Flores}. The term
${\hat I}_0 \left(k/\lambda \right)$ fully contains the infrared
divergence while $C$ and $C_1$ are the finite terms that come along with
it. An alternative approach to extract the infrared divergence and the
terms $C$ and $C_1$ is the one of Ref.~\cite{Ginsberg}, which was applied
to the RC of DP of $K_{e_3}^\pm$ decays. This procedure was followed in
Ref.~\cite{Flores} to extract the infrared divergence of the
spin-dependent part $d\,\Gamma_B^{(s)}$ of its Eq.~(32), and it can also
be employed to extract the infrared divergence of the spin-independent
part $d\,\Gamma_B^\prime$ without any difficulty. In Appendix C we give
a brief review of this, here we only need the result of Eq.~(56) of
Ref.~\cite{Flores}, which is
\begin{equation}
I_0 \left(E,E_2\right) ={\hat I}_0 \left( k/\lambda \right) + C + C_1 .
\label{eq:e26}
\end{equation}
Here $I_0 \left( E,E_2 \right)$ is the infrared-divergent integral given
by Eq.~(\ref{eq:eC6}) of Appendix C. With this equation and the
Eqs.~(36)-(38) of Ref.~\cite{Flores} for $C_2$, $d\,\Gamma_B^{a}$, and
$d\,\Gamma_B^{b}$, respectively, we can express the $\,d\,
\Gamma_B^\prime$ of Eq.~(\ref{eq:e23}), with some minor
rearrangements, in the compact form
\begin{equation}
d\,\Gamma_B^\prime =\frac{\alpha}{\pi}d\Omega^\prime \left\{
A_{1}^\prime I_{0}\left( E,E_{2}\right) +\frac{p_2 l}{4\pi}
\int_{-1}^{1}dx\int_{-1}^{y_{0}}dy\int_{0}^{2\pi}d\phi_{k}\left[ \left|
M^\prime \right|^{2}+\left| M^{\prime \prime} \right|^{2}\right]
\right\} , \label{eq:e27}
\end{equation}
with 
\begin{equation}
\left| M^\prime \right|^2 = \frac{\beta^2 \left(1-x^2 \right)}
{\left(1-\beta x\right)^2} \left[ D_2-\frac{D_1 E+D_2 l x}{D} \right], 
\label{eq:e28}
\end{equation}
and 
\begin{eqnarray}
\left| M^{\prime \prime} \right|^{2} &=&\frac{E_{\nu}}{ED\left( 1-\beta
x\right)} \left[ D_{1}\left( \omega +E\left( 1+\beta x\right) -
\frac{m^{2}}{E \left( 1-\beta x\right)}\right) \right. \nonumber \\
&& \mbox{} + \left. D_{2}{\bf \hat p}_\nu \cdot \left( {\bf l} +
{\bf \hat k} \left( E+\omega \right) -{\bf \hat k}\frac{m^{2}}{E\left(
1-\beta x \right)}\right) \right] . \label{eq:e29}
\end{eqnarray}
In these last three equations, $x= {\bf \hat l}\cdot {\bf \hat k}$ and $y=
{\bf \hat l}\cdot {\bf \hat p}_2$ are the cosines of the polar angles of
${\bf k}$ and ${\bf p}_2$, respectively whereas $\phi_k$ is the azimuthal
angle of ${\bf k}$. Furthermore, $E_\nu = E_\nu^0-\omega$ and
$D=E_{\nu}^0+({\bf p}_2 + {\bf l}) \cdot {\bf \hat k}$. The coefficients
$D_1$ and $D_2$ depend on the leading form factors and they are given in
Eqs.~(\ref {eq:eB11}) and (\ref{eq:eB12}) of Appendix B.
Equation~(\ref{eq:e27}) is also the explicit form of Eq.~(55) of
Ref.~\cite{Tun}.

Once we have $d\,\Gamma_B^\prime$ of Eq.~(\ref{eq:e22}), we can turn
our attention to the spin-dependent part $d\,\Gamma_B^{(s)}$ of this
equation. In order to compute it we shall start from Eq.~(43) of
Ref.~\cite{Flores}, namely,
\begin{equation}
d\,\Gamma_B^{(s)}=d\,\Gamma_B^{\rm I}+d\,\Gamma_B^{\rm II},
\label{eq:e30}
\end{equation}
where $d\,\Gamma_B^{\rm I}$ contains $\sum_{{\rm spins}}\left| M_{a}^{(
s)} \right|^{2}$ and $d\,\Gamma_B^{\rm II}$ contains $\sum_{{\rm spins}}
\left( \left| M_{b}^{(s)} \right|^{2}+2{\rm Re}\left[ M_{a}^{(s)} \right]
\left[ M_{b}^{(s)} \right]^\dagger \right)$. $d\,\Gamma_B^{\rm I}$
contains the infrared-divergent terms as well as many infrared-convergent
ones. $d\,\Gamma_B^{\rm II}$ is infrared-convergent only. To compute
$d\,\Gamma_B^{\rm I}$ we follow the procedure of Ref.~\cite{Ginsberg} to
extract the infrared divergence (see Appendix C). According to this and
using the explicit form of $\sum_{{\rm spins}}\left| M_{a}^{(s)} \right|
^{2}$ given in Eq.~(44) of Ref.~\cite{Flores}, we can write
$d\,\Gamma_B^{\rm I}$ as
\begin{eqnarray}
d\,\Gamma_B^{\rm I} &=&\frac{\alpha}{\pi} d\Omega^\prime \frac{1}{8\pi 
}\lim_{\lambda \rightarrow 0}\int_{\lambda^2}^{\eta_{\max}} d\eta \frac{
d^{3}k}{\omega}\frac{d^{3}p_\nu}{E_\nu}\delta^{4}\left(
p_1 - p_2 - l - p_\nu - k \right) \nonumber \\
&& \mbox{} \times \left[ -D_{3}{\bf \hat s}_1 \cdot {\bf l} E_{\nu
}^{0}+D_{4}{\bf \hat s}_1 \cdot {\bf p}_2 E+D_{4}{\bf \hat s}_1 \cdot
{\bf l} E \right. \nonumber \\
&&\mbox{} + \left. D_{3}{\bf \hat s}_1 \cdot {\bf l}
\omega + D_{4}{\bf \hat s}_1 \cdot {\bf k} E \right]
\times \sum_\epsilon \left( \frac{l \cdot \epsilon}{l \cdot k}-\frac{
p_1 \cdot \epsilon}{p_{1}\cdot k}\right)^{2} . \label{eq:e31}
\end{eqnarray}

The infrared divergence is contained in the first three terms within the
square brackets, the remaining two terms are infrared-convergent. $\eta
=\left(p_\nu+k\right)^2$ is the invariant mass, which in the CM of $A$ is
given by $\eta=2p_2 l \left(y_0-y\right)$. In the TBR, $\eta_{\max}=2p_2 l
\left( y_0+ 1\right)$ and $\eta_{\min} = \lambda^2$, with $\lambda^2
\rightarrow 0$. The coefficients $D_3$ and $D_4$ depend on the leading form
factors and they are given in Eqs.~(\ref{eq:eB13}) and (\ref{eq:eB14}) of
Appendix B. $\epsilon_\mu$ is the polarization four-vector of the real
photon.

In order to express $d\,\Gamma_B^{\rm I}$ in terms of the correlation
${\bf \hat s}_1 \cdot {\bf l}$ we have to transform the scalar products
${\bf \hat s}_1 \cdot {\bf p}_2$ and ${\bf \hat s}_1 \cdot {\bf k}$ of
Eq.~(\ref{eq:e31}) in terms of ${\bf \hat s}_1 \cdot {\bf l}$. This can be
achieved by using the substitutions of Ref.~\cite{Gluck},
\begin{equation}
{\bf \hat s}_1 \cdot {\bf p} \;\; \rightarrow \;\; ({\bf \hat s}_1 \cdot
{\bf \hat l}) \, ({\bf \hat l} \cdot {\bf p}) \qquad ({\bf p} \,= {\bf
p}_2, \; {\bf k}, \;{\bf p}_\nu) . \label{eq:e32}
\end{equation}

In Eq.~(\ref{eq:e31}), with these substitutions, we can separate the
infrared-divergent terms from the infrared-convergent ones as
\begin{eqnarray}
d\,\Gamma_B^{\rm I} &=&\frac{\alpha}{\pi}d\Omega^\prime {\bf \hat s}_1 
\cdot {\bf \hat l}\left\{ B_{2}^\prime \frac{1}{8\pi} \lim_{\lambda
\rightarrow 0}\int_{\lambda^2}^{\eta_{\max}}d\eta \frac{d^{3}k}{\omega}
\frac{d^3 p_\nu}{E_\nu} \delta^4 \left( p_1-p_2-l -p_{\nu
}-k\right) \right. \nonumber \\
&& \times \left[ \frac{2p_1\cdot l}{p_1 \cdot k\,\,l \cdot k}-\frac{m^{2}}{
\left( \l \cdot k\right)^{2}}-\frac{M_{1}^{2}}{\left( p_{1}\cdot k\right)
^{2}}\right]   \nonumber \\
&&\mbox{}+\frac{1}{8\pi}\lim_{\lambda \rightarrow 0} \int_{\lambda
^{2}}^{\eta_{\max}}d\eta \frac{d^{3}k}{\omega}\frac{d^{3}p_\nu}{E_{\nu
}}\delta^{4}\left( p_1 - p_2 - l -p_\nu-k\right)   \nonumber \\
&&\left. \times \left[ D_{3}\omega l +D_{4}{\bf \hat l}\cdot
{\bf \hat k} \,\omega
E-D_{4}\frac{\eta}{2\beta}\right] \frac{\beta^{2}}{\omega^{2}}\frac{
1-\left( {\bf \hat l}\cdot {\bf \hat k} \right)^{2}}{\left( 1-\beta
{\bf \hat l} \cdot {\bf \hat k} \right)^{2}}\right\} . \label{eq:e33}
\end{eqnarray}

Here $B_{2}^\prime$ is given in Eq.~(\ref{eq:e9}). In the first
integral the sum over polarizations indicated in Eq.~(\ref{eq:e31}) was
performed in covariant form, while in the second one it was performed by
using the Coester representation~\cite{Jauch}.

The first integral in Eq.~(\ref{eq:e33}) can be identified with the
divergent integral $I_{0}$ of Eq.~(\ref{eq:eC6}). The second one can be
put in the convenient form of Eq.~(38) of Ref.~\cite{Tun} by performing
the integration over the $\delta $ function and leaving $y$ as the
integration variable instead of $\eta $. In this way, we can express
$d\,\Gamma_B^{\rm I}$ finally as
\begin{eqnarray}
d\,\Gamma_B^{\rm I} &=&\frac{\alpha}{\pi}d\Omega^\prime {\bf \hat s}_1 
\cdot {\bf \hat l} \left\{ B_{2}^\prime I_{0}\left( E,E_{2}\right) +
\frac{p_{2}E}{4\pi}\int_{-1}^{1}dx\int_{-1}^{y_0}dy\int_{0}^{2\pi}d\phi
_{k} \right . \nonumber \\
&&\left. \left[ D_{3}\frac{\beta l}{D}-D_{4}\left( 1-\frac{l x}{D}
\right) \right] \frac{\beta^{2}\left( 1-x^{2}\right)}{\left( 1-\beta
x\right)^{2}}\right\} . \label{eq:e34}
\end{eqnarray}

The other term of Eq.~(\ref{eq:e30}), $d\,\Gamma_B^{\rm II}$, does not
need any calculation. We can take the result of Eq.~(57) of Ref.~\cite
{Flores} and, with only minor changes in the phase space factor, we can
adapt it to our case. After the application of rule~(\ref{eq:e32}) in such
Eq.~(57) is performed, we obtain
\begin{eqnarray}
d\,\Gamma_B^{\rm II} &=&\frac{\alpha}{\pi}d\Omega^\prime {\bf \hat s}_1
\cdot {\bf \hat l} \frac{p_{2}\beta}{4\pi}\int_{-1}^{1}\frac{dx}{ 1-\beta
x}\int_{-1}^{y_{0}}dy\int_{0}^{2\pi}d\phi_k \frac{1}{D} \nonumber \\
&&\times \left\{ D_{3}E_\nu \left[ -l -\left( E+\omega \right) x+
\frac{m^{2}}{E}\frac{x}{1-\beta x}\right] \right.  \nonumber \\
&&\mbox{}+\left. D_{4}\left[ \omega +\left( 1+\beta x\right)
E-\frac{m^{2}}{E }\frac{1}{1-\beta x}\right] \left( p_{2}y+l +\omega
x\right) \right\} . \label{eq:e35}
\end{eqnarray}

At this point we can collect our partial results to get our first main
result, namely the DP of polarized decaying hyperons with radiative
corrections up to order $\alpha$, neglecting terms of order $\alpha q/\pi
M_1$, and restricted to the TBR. It can be set as
\begin{eqnarray}
d\,\Gamma^{\rm TBR} &=&d\,\Gamma_V+d\,\Gamma_{B}^{\rm TBR}  \nonumber \\
& = & d\,\Gamma_V+\left[ d\,\Gamma_{B}^\prime -\left(d\,\Gamma_B^{\rm I}
+ d\,\Gamma_{B}^{\rm II}\right) \right] , \label{eq:e36}
\end{eqnarray}
where $d\,\Gamma_V$ is given in Eq.~(\ref{eq:e7}), $d\,\Gamma_{B}^{\rm
TBR}$ is given in Eq.~(\ref{eq:e22}), and $\,d\Gamma_{B}^\prime$,
$d\, \Gamma_{B}^{\rm I}$ and $d\,\Gamma_{B}^{\rm II}$ are given in
Eqs.~(\ref{eq:e27}), (\ref{eq:e34}), and (\ref{eq:e35}), respectively. The
integrations over the three-momentum of the real photon in these last
three equations are ready to be performed numerically (but they can be
performed analytically also, as we shall see in the next section).

The result of Eq.~(\ref{eq:e36}) can be compared with the corresponding
result of Ref.~\cite{Flores} [the sum of Eqs.~(15), (33), (51), and
(57) of this reference]. We observe that the spin-independent parts
$d\,\Gamma_B^\prime$ are the same, although our present 
$d\,\Gamma_B^\prime$ is expressed in a more compact form, while the
spin-dependent parts are different. The results of Ref.~\cite{Flores} show
that the correlation ${\bf \hat s}_1 \cdot {\bf \hat p}_2$ is mixed with
two others, ${\bf \hat s}_1 \cdot {\bf \hat l}$ and ${\bf \hat s}_1 \cdot
{\bf \hat k}$ [see Eqs.~(51) and (57) of this reference], while our
present results exhibit only the correlation ${\bf \hat s}_1 \cdot {\bf
\hat l}$. If we insisted in transforming the correlations ${\bf \hat s}_1
\cdot {\bf \hat l}$ and ${\bf \hat s}_1 \cdot {\bf \hat k}$ of
Ref.~\cite{Flores} as functions of ${\bf \hat s}_1 \cdot {\bf \hat p}_2$
by using the substitutions of Eq.~(D6) of this reference, the result would
not give the same coefficient of ${\bf \hat s}_1 \cdot {\bf \hat l}$ as
right here above.

\subsection{FBR bremsstrahlung RC and complete RC}

We shall now calculate the bremsstrahlung contribution of the FBR and
afterwards we shall obtain the complete RC to the DP, with the addition of
the TBR and virtual contributions.

The calculation of bremsstrahlung in the FBR is relatively simple because
the events in this region have the same amplitude $M_B$ of Eq.~(\ref
{eq:eA6}) and it is infrared-convergent. We have to change only the upper
limit of the integrals over the variable $y$ of Eqs.~(\ref{eq:e27}), (\ref
{eq:e34}), and (\ref{eq:e35}), which becomes one now, and to change the
previously infrared-divergent integral $I_{0}$ of these equations with
$I_{0F}$ ,
\begin{equation}
I_{0F}=\frac{\theta_{0F}}{2}\ln \left( \frac{y_{0}+1}{y_{0}-1}\right) ,
\label{eq:e37}
\end{equation}
with 
\begin{equation}
\theta_{0F}=4\left( \frac{1}{\beta} {\rm arctanh} \beta -1\right) .
\label{eq:e38}
\end{equation}

$I_{0F}$ is no longer infrared-divergent because in the FBR the photon
has a minimum energy which is nonzero. It can be easily calculated from
Eq.~(\ref{eq:eC4}). The invariant mass $\eta$ of this equation must be now
integrated from a minimum value $\eta_{\min}=2p_2 l \left( y_0-1\right)$
to a maximum value $\eta_{\max}=2p_2 l \left( y_0+1\right)$.

With these changes we can write the differential decay rate corresponding
to the FBR as 
\begin{equation}
d\,\Gamma_B^{\rm FBR}=d\,\Gamma_B^{\prime \, \rm FBR} -
d\,\Gamma_B^{(s) \, \rm FBR} , \label{eq:e39}
\end{equation}
with 
\begin{equation}
d\,\Gamma_B^{\prime \, \rm FBR}=\frac{\alpha}{\pi}d\Omega^\prime
\left\{ A_{1}^\prime I_{0F}\left( E,E_{2}\right) +\frac{p_2 l}{4\pi}
\int_{-1}^{1}dx\int_{-1}^{1}dy\int_0^{2\pi} d\phi_{k}\left[ \left|
M^\prime \right|^{2}+\left| M^{\prime \prime} \right|^{2}\right]
\right\} \label{eq:e40}
\end{equation}
and 
\begin{eqnarray}
d\,\Gamma_B^{(s)\, \rm FBR} &=&\frac{\alpha}{\pi} d\Omega^\prime
{\bf \hat s}_1 \cdot {\bf \hat l} \left\{ B_{2}^\prime I_{0F}\left(
E,E_{2}\right) + \frac{p_2 l}{4\pi} \int_{-1}^{1}dx \int_{-1}^{1}dy 
\int_{0}^{2\pi}d\phi_{k}\left[ \left| M^{\prime \prime
\prime} \right|^{2} + \left| M^{\rm IV} \right|^{2}\right] \right\} .
\label{eq:e41}
\end{eqnarray}

In Eq.~(\ref{eq:e40}), $\left| M^\prime \right|^2$ and $\left|
M^{\prime \prime} \right|^2$ are given in Eqs.~(\ref{eq:e28}) and (\ref
{eq:e29}), respectively. Equation~(\ref{eq:e41}) is the sum of
Eqs.~(\ref{eq:e34}) and (\ref{eq:e35}), after the above changes are
performed. $\left| M^{\prime \prime \prime}\right|^2$ and $\left|
M^{\rm IV} \right|^2$ are
\begin{equation}
\left| M^{\prime \prime \prime} \right|^{2}=\frac{\beta^{2}\left(
1-x^{2}\right)} {\left( 1-\beta x\right)^{2}}\left[ \frac{D_3 l
+D_{4}Ex}{D}-D_{4}\frac{1}{\beta}\right], \label{eq:e42}
\end{equation}
\begin{eqnarray}
\left| M^{\rm IV} \right|^{2} &=&\frac{1}{DE}\left[ D_{3}E_{\nu
}\left( -l -Ex-\omega x+\frac{m^{2}}{E}\frac{x}{1-\beta x}\right) \right.  
\nonumber \\
&&\mbox{}+\left. D_{4}\left( \omega -\frac{m^{2}}{E}\frac{1}{1-\beta x}
+\left( 1+\beta x\right) E\right) \left( p_2 y+l +\omega \right) \right]
. \label{eq:e43}
\end{eqnarray}

Equation~(\ref{eq:e39}) is the FBR-contribution to the RC of the DP. It
can be added to the TBR-contribution to obtain the complete RC to DP of
polarized hyperons within the approximations mentioned before. This
completes our first main result of Eq.~(\ref{eq:e36}) by including the
emission of all real photons allowed by energy-momentum conservation. It
is displayed as
\begin{equation}
d\,\Gamma =d\,\Gamma^{\rm TBR}+d\,\Gamma_{B}^{\rm FBR} , \label{eq:e44}
\end{equation}
with $d\Gamma^{\rm TBR}$ and $d\,\Gamma_{B}^{\rm FBR}$ given in
Eqs.~(\ref{eq:e36}) and (\ref{eq:e39}), respectively. The integrations
over the photon variables are ready to be performed numerically.

Let us close this section by mentioning that all the integrals which arise
in the two regions of the DP can be performed analytically. Because of
this, we can obtain completely analytical results for the RC of DP. We
shall do this in the next section.

\section{Analytical Integrations}

In this section we shall perform analytically the photon three-momentum
integrals contained in Eqs.~(\ref{eq:e27}), (\ref{eq:e34}),
(\ref{eq:e35}), (\ref{eq:e40}), and (\ref{eq:e41}) to obtain an analytical
expression for the RC to the DP restricted to the TBR first and for the
total DP later.

\subsection{TBR Analytical form}

The ${\bf k}$-integrals corresponding to the TBR of the DP are those of
Eqs.~(\ref{eq:e27}), (\ref{eq:e34}), and (\ref{eq:e35}). They can be
performed analytically by following the procedure of Sec. V of
Ref.~\cite{Tun}, where the RC of the DP of unpolarized hyperons were
obtained. Fortunately much of the work has already been advanced. The
integrals that concern us now can be expressed in terms of the functions
$\theta_i$, $i=2,\ldots,9$ given by Eq.~(99) of that reference, and in
terms of the $ \theta_j$, $j=10,\ldots,16$ given by Eq.~(46) of
Ref.~\cite{Juarez}. In this last reference the RC include all the terms of
order $\alpha q/\pi M_1$, which are dropped here. We can express the
analytical form of Eq.~(\ref{eq:e27}) as
\begin{equation}
d\Gamma_{B}^\prime = \frac{\alpha}{\pi}d\Omega^\prime \left[
A_{1}^\prime I_{0}+\left( D_{1}+D_{2}\right) \left( \theta^\prime
+ \theta^{\prime \prime \prime} \right) +D_{2}\left( \theta^{\prime
\prime} + \theta^{\rm IV} \right) \right] . \label{eq:e45}
\end{equation}

This equation is equivalent to $dw_{B}$ of Eq.~(92) of Ref.~\cite{Tun}.
Similarly, the expressions for $d\Gamma_B^{\rm I}$ and $d\Gamma_B^{\rm II}$
of Eqs.~(\ref{eq:e34}) and (\ref{eq:e35}) become 
\begin{equation}
d\Gamma_{B}^{\rm I}=\frac{\alpha}{\pi}d\Omega^\prime \,{\bf \hat s}_1
\cdot {\bf \hat l} \left[ B_{2}^\prime I_{0}+D_{3}\rho_{1}^{\ell} 
+ D_{4}\rho_{2}^{\ell}\right] , \label{eq:e46}
\end{equation}
\begin{equation}
d\Gamma_{B}^{\rm II}=\frac{\alpha}{\pi}d\Omega^\prime \,{\bf \hat s}_1
\cdot {\bf \hat l} \left[ D_{3}\rho_{3}^{\ell}+D_{4}\rho_{4}^\ell \right]
. \label{eq:e47}
\end{equation}

In Eq.~(\ref{eq:e45}) the $\theta_i$ functions are contained in the functions
$\theta^\prime$, $\theta^{\prime \prime}$, $\theta^{\prime \prime
\prime}$ and $\theta^{\rm IV}$ given by Eqs.~(85), (86), (90), and
(91) of Ref.~\cite{Tun}, respectively. The functions $\rho_i^\ell$,
$i=1,\ldots,4$ in Eqs.~(\ref{eq:e46})-(\ref{eq:e47}) can be expressed as
\begin{eqnarray}
\rho_{1}^{\ell} &=&\frac{p_2 l^{2}}{2}\left[ \left( \beta^{2}-1\right)
\theta_{2}+2\theta_{3}-\theta_{4}\right] , \label{eq:e48} \\
& &  \nonumber \\
\rho_{2}^{\ell} &=&\frac{p_{2}E^{2}}{2}\left[ -\frac{2}{E}\theta_0
+ \left(\beta^{2}-1\right) \theta_{2}-\left( \beta^{2}-3\right)
\theta_{3}-2\theta_{4}-\beta \theta_{5}\right] , \label{eq:e49} \\
&& \nonumber \\
\rho_{3}^{\ell} &=&\frac{p_{2}}{2}\left\{ E\left( E+E_{\nu}^{0}\right)
\left( 1-\beta^{2}\right) \theta_{2}-\left[ \left(3 - \beta^{2}\right)
\frac{E^{2}}{2}+EE_{\nu}^{0}\right] \theta_{3}\right.  \nonumber \\
&&\mbox{} +\frac{1}{2}E^{2}\left(1 + \beta^{2}\right)
\theta_{4}-\frac{l}{2} \left( E+2E_{\nu}^{0}\right)
\theta_{5}-\frac{m^{2}}{2E}\theta_{6} \nonumber \\
&& \mbox{} + \left. \frac{1}{2}\left( 2E-E_{\nu}^{0}\right)
\theta_{7}-\frac{1}{2} \left( E-E_{\nu}^{0}\right)
\theta_{8}+\frac{1}{4}\theta_9-\frac{3}{2} l^2
\theta_{10}-\frac{1}{4}\theta_{15}\right\} , \label{eq:e50} \\
&& \nonumber \\
\rho_{4}^{\ell} &=&\frac{p_{2}}{2}\left\{ m^{2}\left[ 2-\beta^{2}+\frac{
E_{\nu}^{0}}{E}\right] \theta_{2}+\left[ -\frac{7}{2}m^{2}+p_2 l y_0
\right] \theta_3 \right.  \nonumber \\
&& \mbox{} +\left[ \left(3 - \beta^{2} \right) \frac{E^2}{2}-EE_\nu^0-p_2
l y_{0}\right] \theta_{4}-\left[ \frac{1}{2}\beta E^{2}+2 l E_\nu^0
\right] \theta_{5} \nonumber \\
&&\mbox{} -\frac{m^{2}}{2E}\theta_{6}+\frac{1}{2}\left( 3E+p_{2}\beta
y_{0}\right) \theta_{7}-E\theta_{8}+\frac{1}{4}\theta_{9}-\frac{5}{2}
\l^2\theta_{10} \nonumber \\
&& \mbox{} - \left. p_2 l \left( 1-\beta^{2}\right) \theta_{11}+2p_2 l
\theta_{12}-p_2 l \theta_{13}-\frac{l}{2}\theta_{14}-\frac{1}{4}
\theta_{15}-\frac{1}{4E}\theta_{16}\right\} . \label{eq:e51}
\end{eqnarray}

With Eqs.~(\ref{eq:e45})-(\ref{eq:e47}) all the integrals over ${\bf k}$
in Eqs.~(\ref{eq:e27}), (\ref{eq:e34}), and (\ref{eq:e35}) have been
expressed in an analytical form. We can obtain now the bremsstrahlung
differential decay rate $d\Gamma_B^{\rm TBR} $ of decaying polarized
hyperons with the photon integrals expressed analytically. Substituting in
Eq.~(\ref{eq:e22}) the analytical forms of $d\Gamma_B^\prime$
Eq.~(\ref{eq:e45}) and of $d\Gamma_{B}^{(s)}$, which is the sum of
Eqs.~(\ref{eq:e46}) and (\ref{eq:e47}), we obtain the analytical form of
$d\Gamma_{B}^{\rm TBR}$
\begin{eqnarray}
d\Gamma_{B}^{\rm TBR} &=&\frac{\alpha}{\pi}d\Omega^\prime \left\{
A_{1}^\prime I_{0}+\left( D_{1}+D_{2}\right) \left( \theta^\prime 
+\theta^{\prime \prime \prime} \right) +D_{2}\left( \theta^{\prime
\prime} +\theta^{\rm IV} \right) \right.  \nonumber \\
&& \mbox{} - \left. {\bf \hat s}_1 \cdot {\bf \hat l} \left[
B_{2}^\prime I_{0}+D_3 \left(\rho_1^\ell + \rho_3^\ell \right) + D_4
\left(\rho_2^\ell+\rho_4^\ell \right) \right] \right\} . \label{eq:e52}
\end{eqnarray}

We are now in a position to obtain our second main result in this paper:
the analytical RC to the DP of decaying polarized hyperons to order
$\alpha$ and neglecting terms of order $\alpha q/\pi M_1$. This result
comes from the addition of the virtual RC, $d\Gamma_V$ of
Eq.~(\ref{eq:e7}), and of $ d\Gamma_B^{\rm TBR}$ of Eq.~(\ref{eq:e52}). It
can be put compactly as
\begin{equation}
d\Gamma^{\rm TBR} = \frac{G_V^{2}}{2} \frac{dE_{2} \, dE \,
d\Omega_{\ell}}{\left( 2\pi \right)^{4}}2M_{1}\left\{
A_{0}^\prime +\frac{\alpha}{\pi}\Phi_1- {\bf \hat s}_1 \cdot {\bf
\hat l} \left[ B_{0}^{\prime \prime}+ \frac{\alpha}{
\pi}\Phi_{2}^{\ell}\right] \right\} . \label{eq:e53}
\end{equation}

Here $A_{0}^\prime$ and $B_{0}^{\prime \prime}$ are the same as
Eqs.~(\ref{eq:eB1}) and~(\ref{eq:e8}), respectively. $\Phi_{1}$ and 
$\Phi_{2}^{\ell}$ are
\begin{eqnarray}
\Phi_{1} &=&A_{1}^\prime \left( \phi +I_{0}\right) +A_{1}^{\prime
\prime} \phi^\prime +\left( D_1+D_2 \right) \left(\theta^\prime +
\theta^{\prime \prime \prime} \right) + D_2 \left( \theta^{\prime
\prime} + \theta^{\rm IV} \right) , \label{eq:e54} \\
&&  \nonumber \\
\Phi_2^\ell &=& B_2^\prime \left(\phi + I_0 \right) + B_2^{\prime
\prime} \phi^\prime + D_3 \left(\rho_1^\ell + \rho_3^\ell \right)
+D_{4}\left( \rho_{2}^{\ell}+\rho_{4}^{\ell}\right) . \label{eq:e55}
\end{eqnarray}

The coefficients $A_1^\prime$, $A_1^{\prime \prime}$, $B_2^\prime$,
and $B_2^{\prime \prime}$ are given in Eqs.~(\ref{eq:eB2}),
(\ref{eq:eB3}), (\ref{eq:e9}) and (\ref{eq:e10}), respectively. $D_i$,
$i=1,\ldots,4$ are given in Eqs.~(\ref{eq:eB11})-(\ref{eq:eB14}). The
functions $\phi$, $\phi^\prime$, and $I_0$ appear in
Eqs.~(\ref{eq:e11}), (\ref{eq:e12}), and (\ref{eq:eC6}), respectively. The
new model-independent functions $\rho_i^\ell$, $i=1,\ldots,4$ were given
in Eqs.~(\ref{eq:e48})-(\ref{eq:e51}). The sums $\theta^\prime +
\theta^{\prime \prime \prime}$ and $\theta^{\prime \prime} +
\theta^{\rm IV}$ appear explicitly in Eqs.~(93) and (94) of
Ref.~\cite{Tun}, respectively. We have corrected a misprint in that
Eq.~(93). Its fourth term has to be $\frac{l}{2} \theta_5$ rather than
$-\frac{l}{2} \theta_5$. For completeness, we reproduce these two sums in
Appendix B [see Eqs.~(\ref{eq:eB38}) and (\ref{eq:eB39})].

Because the infrared divergence, which appears in the virtual part 
$d\Gamma_V$ through the function $\phi$, cancels out with its
bremsstrahlung counterpart, which appears in $I_0$, the sum $\phi +I_0$ is
no longer infrared-divergent and, therefore, $d\Gamma^{\rm TBR}$ of
Eq.~(\ref {eq:e53}) is infrared-convergent.

$d\Gamma^{\rm TBR}$ is the result corresponding to $d\Gamma$ of Eq.~(101)
in Ref.~\cite{Flores}. Comparing both results, we can observe that the
spin-independent parts are the same, but the spin-dependent parts show
important differences. In Eq.~(101) of Ref.~\cite{Flores} we have, within
the square brackets that accompany the correlation ${\bf \hat s}_1 \cdot
{\bf \hat p}_2$, the terms $A_0^{\prime \prime}$ and $\Phi_2$, while in
Eq.~(\ref{eq:e53}) the corresponding terms accompanying ${\bf \hat s}_1
\cdot {\bf \hat l}$ are $B_0^{\prime \prime}$ and $\Phi_2^\ell$. They
are different. In fact, we may notice that in $\Phi_2^\ell$ of
Eq.~(\ref{eq:e55}) only the $\theta_i$-functions, $i=2,\ldots,16$
appear, while in $\Phi_2$ of Eq.~(103) of Ref.~\cite{Flores} also the
$\eta$-functions appear [see Eqs.~(91)-(95) of this reference]. Because
of this, the RC to ${\bf \hat s}_1 \cdot {\bf \hat l}$ and ${\bf \hat s}_1
\cdot {\bf \hat p}_2$ correlations are quite different.

\subsection{FBR analytical form}

The ${\bf k}$-integrals of the FBR are those contained in
Eqs.~(\ref{eq:e40}) and (\ref{eq:e41}) and they can be performed
analytically, too. Because $d\Gamma_B^{\prime \, \rm FBR}$ of
Eq.~(\ref{eq:e40}) has the same form as the corresponding
$d\Gamma_B^\prime$ of Eq.~(\ref {eq:e27}) of the TBR, we can follow the
same procedure of the Sec. V of Ref.~\cite{Tun} to calculate the
analytical form of this $d\Gamma_B^{\prime \, \rm FBR}$. The result has
the same structure as $d\Gamma_B^\prime$ of Eq.~(\ref{eq:e45}),
\begin{equation}
d\Gamma_B^{\prime \,\rm FBR}=\frac{\alpha}{\pi}d\Omega^\prime
\left[ A_1^\prime I_{0F}+\left( D_{1}+D_{2}\right) \left( \theta
_{F}^\prime+\theta_F^{\prime \prime \prime} \right) +D_{2}\left(
\theta_F^{\prime \prime}+\theta_F^{\rm IV} \right) \right],
\label{eq:e56}
\end{equation}
with 
\begin{eqnarray}
\theta_F^\prime +\theta_F^{\prime \prime \prime} &=&\frac{
p_2 l}{2}\left[ -E_\nu^{0}\left( 1-\beta^{2}\right) \theta
_{2F}+\left( E_\nu^0-\frac{1+\beta^{2}}{2}E\right) \theta_{3F}+\frac{E
}{2}\theta_{4F} \right. \nonumber \\
&& \mbox{} +\left. \frac{l}{2}\theta_{5F}+\frac{1-\beta^{2}}{2}\theta
_{6F}-\frac{2E-E_\nu^0}{2E}\theta_{7F}+\frac{1}{2}\theta_{8F}-\frac{1
}{4E}\theta_{9F}\right] , \label{eq:e57}
\end{eqnarray}
and 
\begin{equation}
\theta_F^{\prime \prime} + \theta_F^{\rm IV} = \frac{p_2 l}{2}
\left[ \theta_{0F} - \left( E+E_\nu^0 + \beta p_2 y_0 \right) \theta_{3F}
+ \left(E_\nu^0 + E\right) \theta_{4F} + l \theta_{5F} \right]
. \label{eq:e58}
\end{equation}

These last two equations also have the same structure as
$\theta^\prime + \theta^{\prime \prime \prime}$ and
$\theta^{\prime \prime} + \theta^{\rm IV}$, which appear in the
analytical form of $d\Gamma_B^\prime$ of Eq.~(\ref{eq:e45}). The
difference between $d\Gamma_{B}^{\prime \, \rm FBR}$ and
$d\Gamma_B^\prime$ of the TBR lies in the integrals $I_{0}$, $I_{0F}$
and in the functions $\theta_i$. In the FBR these functions change to
$\theta_{iF}$ because the variable $y$ must be integrated from $-1$ to $1$
and not from $-1$ to $y_{0}$ as in the TBR. The new set $\left\{
\theta_{iF} \right\} ,i=2,\ldots,16$ is explicitly given in Appendix B and
$\theta_{0F}$ is given in Eq.~(\ref{eq:e38}). We can compare with the
$\theta_{i}^{T}$ of Ref.~\cite{S. R. Juarez} where the RC of DP for the
FBR were calculated up to order $\alpha q/\pi M_1$. We cannot take readily
the result of this reference because, according to our approximations, we
would have to neglect all the terms of order $\alpha q/\pi M_{1}$ in that
result to obtain ours. We find the procedure of Ref.~\cite{Tun} more
adequate and straightforward for our purposes. However, in order to check
our results we have reproduced the Table I of Ref.~\cite{S. R. Juarez}. Our
numerical evaluations coincide very well within the approximation of
neglecting terms of order $\alpha q/\pi M_1$; we shall not display here
this numerical evaluation.

In a similar way, we can see that the spin-dependent part of
$d\Gamma_B^{\rm FBR}$ has the same structure as the corresponding
spin-dependent part of $d\Gamma_B^{\rm TBR}$. Thus, we get
\begin{equation}
d\Gamma_{B}^{(s) \, \rm FBR}=\frac{\alpha}{\pi} d\Omega^\prime
\,\, {\bf \hat s}_1 \cdot {\bf \hat l} \left[ B_{2}^\prime
I_{0F}+D_{3}\left( \rho_{1F}^{\;\ell} + \rho_{3F}^{\;\ell} \right)
+D_{4}\left( \rho_{2F}^{\;\ell}+\rho_{4F}^{\;\ell}\right) \right] \, ,
\label{eq:e59}
\end{equation}
with the functions $\rho_{iF}^{\;\ell}$, $i=1,\ldots,4$ having the
same structure as the previous $\rho_{i}^{\ell}$ of
Eqs.~(\ref{eq:e48})-(\ref{eq:e51}). Explicitly, they are 
\begin{eqnarray}
\rho_{1F} &=&\frac{p_2 l^2}{2}\left[ \left( \beta^{2}-1\right)
\theta_{2F}+2\theta_{3F}-\theta_{4F}\right] , \label{eq:e60} \\
&&  \nonumber \\
\rho_{2F} &=&\frac{p_{2}E^{2}}{2}\left[ -\frac{2}{E}\theta_{0F}+\left(
\beta^{2}-1\right) \theta_{2F}-\left( \beta^{2}-3\right) \theta
_{3F}-2\theta_{4F}-\beta \theta_{5F}\right] , \label{eq:e61} \\
& &  \nonumber \\
\rho_{3F} &=&\frac{p_{2}}{2}\left\{ E\left( E+E_{\nu}^{0}\right) \left(
1-\beta^{2}\right) \theta_{2F}-\left[ \left( 3-\beta^{2}\right) \frac{ E^{2}
}{2}+EE_{\nu}^{0}\right] \theta_{3F}\right. \nonumber \\
&& \mbox{} + \frac{1}{2}E^{2}\left(1 + \beta^{2} \right)
\theta_{4F}-\frac{l}{2} \left( E+2E_{\nu}^{0}\right)
\theta_{5F}-\frac{m^{2}}{2E}\theta_{6F} \nonumber \\
&&\mbox{} + \left. \frac{1}{2}\left(
2E-E_{\nu}^{0}\right) \theta_{7F}-\frac{1}{2}
\left( E-E_{\nu}^{0}\right) \theta_{8F}+\frac{1}{4}\theta_{9F}- \frac{3}{2}
l^2 \theta_{10F}-\frac{1}{4}\theta_{15F}\right\} , \label{eq:e62} \\
&& \nonumber \\
\rho_{4F} &=&\frac{p_{2}}{2}\left\{ m^{2}\left[ 2-\beta^{2}+\frac{E_{\nu
}^{0}}{E}\right] \theta_{2F}+\left[ -\frac{7}{2}m^{2}+p_2 l y_0 \right]
\theta_{3F}\right.  \nonumber \\
&&+\left[ \left( 3-\beta^{2}\right) \frac{E^{2}}{2}-EE_{\nu}^{0}-p_2 l
y_{0}\right] \theta_{4F}-\left[ \frac{1}{2}\beta E^{2}+2 l E_{\nu
}^{0}\right] \theta_{5F}  \nonumber \\
&& \mbox{} - \frac{m^{2}}{2E}\theta_{6F}+\frac{1}{2}\left( 3E+p_{2}\beta
y_{0}\right) \theta_{7F}-E\theta_{8F}+\frac{1}{4}\theta_{9F}-\frac{5}{2} l
^{2}\theta_{10F}  \nonumber \\
&&\mbox{} - \left. p_2 l \left( 1-\beta^{2}\right) \theta_{11F}+2p_2 l
\theta_{12F}-p_2 l \theta_{13F}-\frac{l}{2}\theta_{14F}- \frac{1}{4}
\theta_{15F}-\frac{1}{4E}\theta_{16F}\right\} . \label{eq:e63}
\end{eqnarray}

From Eqs.~(\ref{eq:e56}) and (\ref{eq:e59}) we obtain the analytical
bremsstrahlung differential decay rate $d\Gamma_B^{\rm FBR}$ of decaying
polarized hyperons corresponding to Eq.~(\ref{eq:e39}), 
\begin{equation}
d\Gamma_B^{\rm FBR}=\frac{\alpha}{\pi}d\Omega^\prime \left[
\Phi_{1F}-{\bf \hat s}_1 \cdot {\bf \hat l} \; \Phi_{2F}^\ell
\right] , \label{eq:e64}
\end{equation}
with 
\begin{eqnarray}
\Phi_{1F} &=&A_{1}^\prime I_{0F}+\left( D_{1}+D_{2}\right) \left(
\theta_F^\prime + \theta_F^{\prime \prime \prime}\right) +D_{2}\left(
\theta_F^{\prime \prime} + \theta_F^{\rm IV} \right) ,
\label{eq:e65} \\ 
&& \nonumber \\
\Phi_{2F}^{\;\ell} &=&B_{2}^\prime I_{0F}+D_{3}\left(
\rho_{1F}^{\;\ell}+\rho_{3F}^{\;\ell}\right) +D_{4}\left(
\rho_{2F}^{\;\ell }+\rho_{4F}^{\;\ell}\right) . \label{eq:e66}
\end{eqnarray}

At this point we complete our second main result. The addition of
$d\Gamma_B^{\rm FBR}$ of Eq.~(\ref{eq:e64}) and $d\Gamma^{\rm
TBR}$ of Eq.~(\ref{eq:e53}) gives us the complete analytical RC to the DP
of decaying polarized hyperons to order $\alpha$ and neglecting terms of
order $\alpha q/\pi M_{1}$. This complete result can be expressed
compactly as 
\begin{eqnarray} 
d\Gamma_{\rm TOT} &=&\frac{G_V^{\,\,2}}{2} \frac{dE_{2}\,dE\,
d\Omega_\ell}{ \left( 2\pi \right)^{4}} 2M_{1}\left\{ A_{0}^\prime
+ \frac{\alpha}{\pi} \left( \Phi_1+\Phi_{1F}\right) - {\bf \hat s}_1
\cdot {\bf \hat l}\left[ B_{0}^{\prime \prime} + \frac{\alpha}{\pi}
\left(\Phi_2^\ell + \Phi_{2F}^\ell \right) \right] \right\} \, .
\label{eq:e67}
\end{eqnarray}
Here $A_0^\prime$, $B_{0}^{\prime \prime}$, $\Phi_1$, $\Phi_{1F}$,
$\Phi_{2}^\ell$, and $\Phi_{2F}^\ell$ are given in Eqs.~(\ref{eq:eB1}),
(\ref{eq:e8}), (\ref{eq:e54}), (\ref{eq:e65}), (\ref {eq:e55}), and
(\ref{eq:e66}), respectively.

From Eq.~(\ref{eq:e67}) we can obtain easily Eq.~(\ref{eq:e53}) to the RC
of DP with the TBR only by dropping $\Phi_{1F}$ and $\Phi_{2F}^\ell$. It
is this Eq.~(\ref{eq:e67}) which must be used to obtain, in HSD, totally
integrated observables, such as the spin-asymmetry coefficient of the
charged lepton. We shall calculate this asymmetry coefficient in the next
section, allowing for the possibility that real photon emission be
discriminated experimentally via energy-momentum conservation.

\section{Spin asymmetry coefficient $\alpha_\ell$}

In this section we shall obtain the RC to the spin-asymmetry coefficient
of the charged lepton $\alpha_\ell$. We shall consider the two cases
discussed all along, namely, that bremsstrahlung photons not be
discriminated at all or that directly or indirectly the photons belonging
to the FBR be eliminated from the experimental analysis. We will discuss
the former case first and afterwards we will discuss the latter case. As
we shall see in the numerical evaluation of the next section, an
appreciable difference can be observed between these two cases.

$\alpha_\ell$ can be calculated from the total DP of Eq.~(\ref{eq:e67}).
This equation can be used to get the quantities $N^{\pm}$ which appear in
the definition of $\alpha_\ell$, 
\begin{equation}
\alpha_\ell= 2 \frac{N^{+}-N^{-}}{N^{+}+N^{-}} . \label{eq:e68}
\end{equation}

Here $N^+$ $(N^-)$ denotes the number of the emitted charged leptons with
momenta in the forward (backward) hemisphere with respect to the
polarization of the decaying hyperon. With those numbers calculated, we
may express $\alpha_\ell$ as 
\begin{equation}
\alpha_\ell^{\rm T}=-\frac{B_{2}^{\;\ell}+\left( \alpha \,/\,\pi \right)
\left( a_{2}^{\;\ell}+a_{2F}^{\;\ell}\right)}{B_{1}+\left( \alpha \,/\,\pi
\right) \left( a_{1}+a_{1F}\right)} . \label{eq:e69} 
\end{equation}

Here 
\begin{eqnarray}
B_2^\ell &=& \int_m^{E_m} \int_{E_2^-}^{E_2^+}B_{0}^{\prime \prime} 
dE_{2}dE , \label{eq:e70} \\
& & \nonumber \\
a_{2}^\ell &=&\int_{m}^{E_{m}}\int_{E_{2}^{-}}^{E_{2}^{+}} 
\Phi_{2}^{\,\,\,\ell}dE_{2}\,dE , \label{eq:e71} \\
& &  \nonumber \\
a_{2F}^{\ell} &=&\int_{m}^{E_{B}}\int_{M_{2}}^{E_{2}^{-}} \Phi_{2F}^{\ell}
dE_{2}\,dE , \label{eq:e72} \\
& &  \nonumber \\
B_{1} &=&\int_{m}^{E_{m}}\int_{E_{2}^{-}}^{E_{2}^{+}}A_{0}^\prime
dE_{2}\,dE , \label{eq:e73} \\
& &  \nonumber \\
a_{1} &=&\int_{m}^{E_{m}}\int_{E_{2}^{-}}^{E_{2}^{+}}\Phi_1 dE_{2}\,dE ,
\label{eq:e74} \\
& &  \nonumber \\
a_{1F} &=&\int_{m}^{E_{B}}\int_{M_{2}}^{E_{2}^{-}}\Phi_{1F}\,dE_{2}\,dE .
\label{eq:e75}
\end{eqnarray}
In these integrals, $B_0^{\prime \prime}$, $\Phi_2^{\ell}$,
$\Phi_{2F}^{\ell}$, $A_0^\prime$, $\Phi_1$, and $\Phi_{1F}$ are given
in Eqs.~(\ref{eq:e8}), (\ref{eq:e55}), (\ref{eq:e66}), (\ref {eq:eB1}),
(\ref{eq:e54}), and (\ref{eq:e65}), respectively.

In Eq.~(\ref{eq:e69}) we have attached an upper index T to denote that
the asymmetry coefficient includes the total DP of the real photons. The
contributions of the TBR to the RC of $\alpha_\ell$ are given by the terms
$a_2^\ell$ and $a_{1}$, while the contributions of the FBR are given by
the terms $a_{2F}^{\;\ell}$ and $a_{1F}$. We can now rewrite 
$\alpha_\ell^{\rm T}$ to comply with our approximations, {\it i.e.}, in
such a way that only the terms of order $\alpha$, neglecting terms of
order $\alpha q/\pi M_1$, appear. The corresponding expression is
\begin{equation}
\alpha_\ell^{\rm T}=\alpha_{0}^{\ell}\left[ 1+\frac{\alpha}{\pi}\left(
\frac{ a_{2}^{\ell}+a_{2F}^{\ell}}{B_{2}^{\ell}\left( 0\right)}-
\frac{a_{1}+a_{1F}}{B_{1}\left( 0\right)}\right) \right] , \label{eq:e76}
\end{equation}
where $\alpha_0^\ell$ is the spin-asymmetry coefficient of the charged
lepton without RC. It is obtained from Eq.~(\ref{eq:e69}) by dropping the
terms proportional to $\alpha$, namely,
\begin{equation}
\alpha_{0}^{\ell}=-\frac{B_{2}^{\ell}}{B_{1}} . \label{eq:e77}
\end{equation}

$B_{2}^{\ell}\left( 0\right) $ and $B_{1}\left( 0\right) $ in the
denominators of Eq.~(\ref{eq:e76}) are the zero-order $q/M_{1}$
approximations of the $B_{2}^{\ell}$ of Eq.~(\ref{eq:e70}) and of the
$B_{1}$ of Eq.~(\ref{eq:e73}), respectively. Explicitly they are,
\begin{eqnarray}
B_{2}^{\ell}\left( 0\right)
&=&\int_{m}^{E_{m}}\int_{E_{2}^{-}}^{E_{2}^{+}}B_{2}^\prime \, dE_{2} \, dE
, \label{eq:e78} \\
&&  \nonumber \\
B_{1}\left( 0\right) & = & \int_{m}^{E_{m}} \int_{E_{2}^{-}}^{E_{2}^{+}}
A_1^\prime \, dE_{2} \, dE , \label{eq:e79}
\end{eqnarray}
with $B_{2}^\prime$ and $A_{1}^\prime$ given in Eqs.~(\ref{eq:e9})
and (\ref{eq:eB2}), respectively.

The coefficient for $\alpha_{\ell}$ when only the TBR of the DP is allowed
can be easily obtained now. All that has to be done is to drop
$a_{2F}^{\ell}$ and $a_{1F}$ from Eq.~(\ref{eq:e76}) so that
\begin{equation}
\alpha_{\ell}^{\rm R}=\alpha_{0}^{\ell}\left[ 1+\frac{\alpha}{\pi}\left(
\frac{ a_{2}^{\ell}}{B_{2}^{\ell}\left( 0\right)}-\frac{a_{1}}{B_{1}\left(
0\right)} \right) \right] . \label{eq:e80}
\end{equation}
We attached an upper index R to denote that the bremsstrahlung correction
is restricted to the TBR.

In Ref.~\cite{Flores} we only calculated the emitted baryon
asymmetry-coefficient $\alpha_{B}$ corresponding to $\alpha_\ell^{\rm R}$.
In this reference it was assumed that the FBR photons were always
discriminated. The contribution of these photons should be calculated and
added to the results of this reference in order to get an $\alpha_{B}$
corresponding to the above $\alpha_{\ell}^{\rm T}$.

In the next section we shall display numerical evaluations that will allow
us to compare our results with others available in the literature and,
also, to appreciate the relevance of discriminating or not FBR photons.

\section{Numerical result}

In order to compare the coefficients $\alpha_{\ell}^{\rm R}$ of the TBR of
the DP and $\alpha_{\ell}^{\rm T}$ of the total DP, we shall make
numerical evaluations of them for several decays. These results will
enable us to establish the relevance of the difference between
$\alpha_{\ell}^{\rm R}$ and $\alpha_{\ell}^{\rm T}$ in the study of HSD.
We shall also compare them with other results reported in the literature.
In Table I we give the values of the form factors used in the numerical
evaluation of the coefficients $ \alpha_{\ell}^{\rm R}$ and
$\alpha_{\ell}^{\rm T}$ for the decays $n\rightarrow pe \overline{\nu}$,
$\Lambda \rightarrow pe\overline{\nu}$, $\Sigma^- \rightarrow
ne\overline{\nu}$, $\Sigma^-\rightarrow \Lambda e \overline{\nu}$,
$\Sigma^+\rightarrow \Lambda e^{+}\nu$, $\Xi^- \rightarrow \Lambda e
\overline{\nu}$, $\Xi^- \rightarrow \Sigma^0 e\bar{\nu}$, $\Xi^0
\rightarrow \Sigma^+ e\overline{\nu}$, and $\Lambda_c^+ \rightarrow
\Lambda e^+ \nu$. For this last decay we take the form factors of
Ref.~\cite{Perez}. The sign of the form factor $g_{1}$ must be changed
when the charged lepton is positive \cite{Martinez}. In the radiatively
uncorrected amplitudes the $q^{2}$-dependence of the form factor was
neglected along with the contributions arising from $f_3$, $g_2$, and $g_3$
as was done in other calculations in the literature.

To evaluate $\alpha_{\ell}^{\rm R}$ we use Eq.~(\ref{eq:e80}) and for $
\alpha_\ell^{\rm T}$ we use Eq.~(\ref{eq:e76}). These equations involve
the double integration over the energies $E$ and $E_{2}$. At this point is
convenient to mention a technical aspect that we have to deal with in
calculating the integrals
\begin{equation}
\int_{E_2^-}^{E_2^+} \ln \left(y_0+1\right) dE_2 , \label{eq:e81}
\end{equation}
and 
\begin{equation}
\int_{M_{2}}^{E_{2}^{+}}\ln \left( \frac{y_{0}+1}{y_{0}-1}\right) dE_{2} ,
\label{eq:e82}
\end{equation}
which are contained in the infrared-divergent integral $I_{0}$ of
Eq.~(\ref {eq:eC6}) and in the integral $I_{0F}$ of Eq.~(\ref{eq:e37}),
respectively. The integrals of Eqs.~(\ref{eq:e81}) and
(\ref{eq:e82}) diverge when $y_{0}\rightarrow \pm 1$, that is when
$E_{2}=E_{2}^{\pm}$. However, it turns out that the result is finite as was
shown in Ref.~\cite{Castro}. To calculate it we can either follow this
approach and implement it in the program to evaluate the $ \alpha_{\ell}$'s
or we may neglect the points where $y_{0}\rightarrow \pm 1$ . This last is
equivalent to leaving out the boundaries of the TBR and the FBR of the
DP. The numerical difference between these two alternatives is
negligible. Here we follow the second alternative.

In Table II we display our numerical results for the radiative corrections
of the asymmetry coefficients $\alpha_{\ell}^{\rm R}$ and
$\alpha_{\ell}^{\rm T}$. We compute them by taking the percentage
differences (that is, we multiply by 100)
\begin{equation}
\delta \alpha_{\ell}^{\rm R,T}=\alpha_{\ell}^{\rm R,T}-\alpha_0^\ell , 
\label{eq:e83}
\end{equation}
where $\alpha_{0}^{\ell}$ is the uncorrected spin-asymmetry coefficient of
the charged lepton, Eq.~(\ref{eq:e77}).

In the second column of Table II we display the $\delta \alpha^{\rm R}$
corresponding to the TBR of the DP, in the third column the $\delta
\alpha^{\rm T}$ corresponding to the complete DP are given, in the fourth
column we give the results for $\delta \alpha$ obtained from Eq.~(23) and
Table I of page 58 of Ref.~\cite{Garcia} and, finally, in the fifth column
we give the two values reported in Ref.~\cite{Gluck}.

From this Table II we see that there is a very good agreement between our
$\delta \alpha^{\rm T}$ and the $\delta \alpha$ of Refs.~\cite{Garcia} and
\cite{Gluck}. In both these references the FBR was included. The inclusion
or exclusion of the FBR is appreciable, as can be seen by comparing the
second and third columns, except for the decays $n\rightarrow
pe\overline{\nu}$ and $\Sigma^-\rightarrow ne\overline{\nu}$. In several
instances the inclusion of the FBR contributions reduces the total
radiative corrections, even to the point of making them negligibly small.
It may be that the values in the second column are one order of magnitude
larger than the corresponding ones in the third column. Therefore, in
general, there is an important difference between $\alpha_{\ell}^{\rm R}$
and $\alpha_\ell^{\rm T}$. Experiments of HSD where no provision to detect
the photon has been made must use $\alpha_\ell^{\rm T}$ to study such
decays. But, if the photon is discriminated eliminating the FBR of the DP
then the adequate coefficient to study HSD is $\alpha_\ell^{\rm R}$.

\section{Conclusions}

In this paper we have obtained the radiative corrections to order $\alpha$
to the Dalitz plot of the semileptonic decays of polarized spin-$1/2$
baryons, neglecting terms of order $\alpha q/\pi M_1$ and higher. Our
main result, which is compactly given by
\begin{equation}
d\Gamma =d\Gamma^{\rm TBR}+d\Gamma_B^{\rm FBR} , \label{eq:e84}
\end{equation}
is presented in two forms: one in which the triple ${\bf k}$-integration
is ready to be performed numerically, Eqs.~(\ref{eq:e36}) and
(\ref{eq:e44}), and another one in which such integration has been
analytically performed, Eqs.~(\ref{eq:e53}) and (\ref{eq:e67}). Since real
photons may be discriminated either directly (by detection) or indirectly
(by energy-momentum conservation) we have split the above two forms to
cover this possibility. In case this photon discrimination takes place,
Eq.~(\ref {eq:e84}) becomes
\begin{equation}
d\Gamma^{\rm TBR}=d\Gamma_V +\left[ d\Gamma_B^\prime-\left(
d\Gamma_B^{\rm I} + d\Gamma_B^{\rm II} \right) \right] , \label{eq:e85}
\end{equation}
where $d\Gamma_V$ is given in Eq.~(\ref{eq:e7}) and it corresponds to the
virtual RC. $d\Gamma_B^\prime$, $d\Gamma_B^{\rm I}$, and $d\Gamma_B^{\rm
II}$ correspond to the bremsstrahlung RC and are given by 
Eqs.~(\ref{eq:e27}), (\ref{eq:e34}), and (\ref{eq:e35}), respectively,
with the photon integrals remaining to be performed numerically. The
infrared divergence and the finite terms that accompany it have been
explicitly and analytically extracted, however. They appear in the $I_{0}$
of $d\Gamma_B^\prime$ and of $d\Gamma_B^{\rm I}$. The analytical result
for Eq.~(\ref{eq:e85}) is given in Eq.~(\ref{eq:e53}).

When real photons are not discriminated what-so-ever, then the DP with the
union of the TBR and the FBR must be used, as in Eq.~(\ref{eq:e84}). $
d\Gamma_B^{\rm FBR}$ is given in Eq.~(\ref{eq:e39}), where the ${\bf
k}$-integration remains to be evaluated numerically, and the analytical
result for Eq.~(\ref{eq:e84}) is given in Eq.~(\ref{eq:e67}).

An important integrated observable is the charged-lepton spin-asymmetry
coefficient $\alpha_{\ell}$. Using the analytical form we obtained the
radiative corrections, through Eq.~(\ref{eq:e67}), to this observable. The
integrations over $E$ and $E_2$ were performed numerically and the results
are displayed in Table II. A systematic behavior of the RC to
$\alpha_\ell$ is observed. The contribution of the FBR bremsstrahlung may
be as important as the RC from the TBR and even of opposite sign, in such
a way that when no photon discrimination takes place the complete RC to
$\alpha_\ell$ may become almost negligible. In this Table we also compare
with results reported in other references. This comparison is
satisfactory.

To our knowledge Eqs.~(\ref{eq:e53}) and (\ref{eq:e67}) for the RC to the
DP allowing for the TBR only and including the FBR are the only analytical
expressions available in the literature. In Ref.~\cite{Gluck} the relevant
variables of the DP are also $E$ and $E_{2}$, but there only numerical
results are given, which are compromised to particular values of the form
factors. Our numerical evaluations coincide very well with the numbers
reported there for the percent RC of the asymmetry coefficient $
\alpha_\ell^{\rm T}$ for the decays $\Sigma^-\rightarrow ne\overline{\nu}$
and $\Lambda \rightarrow pe\overline{\nu}$. We have also evaluated the RC
for the process $\Lambda_c^+ \rightarrow \Lambda e^{+}\nu$. The numbers
obtained are presented in Table II and in this decay they are a good first
approximation, which is already useful given that the experimental
statistics are not yet too high.

Our results are model-independent and are not compromised to any
particular value of the form factors. All the model dependence of
radiative corrections has been absorbed into the $f_{1}$ and $g_{1}$ form
factors in our approximation of neglecting contributions of order $\alpha
q/\pi M_1$. This is indicated by putting a prime on them. For hyperons our
results are reliable up to a precision of around $0.5\%$. This precision
is useful for experiments involving several thousands of events. For high
statistics experiments involving several hundreds of thousands of events
or for decays involving charm as $\Lambda_c^+ \rightarrow \Lambda e^+\nu$
or even heavier quarks our equations provide a good first approximation.
To improve the precision of our formulas it becomes necessary to include
$\alpha q/\pi M_{1}$ contributions. Our results are valid both for neutral
or charged decaying hyperons and whether the emitted positively or
negatively charged lepton is either electron-type or muon-type. To
conclude let us remark that in a Monte Carlo analysis the advantage of the
analytical form is that the triple ${\bf k}$-integration does not have to
be repeated every time the values of $f_1$ and $g_1$, or of $E$ and
$E_{2}$, are changed. This leads to a considerable saving of computer time.

\acknowledgements

The authors are grateful to Consejo Nacional de Ciencia y Tecnolog{\'\i}a
(M\'{e}xico) for partial support. A. M. gratefully acknowledges partial
support by Comisi\'on de Operaci\'on y Fomento de Actividades Acad\'emicas
(Instituto Polit\'{e}cnico Nacional). R.F.M. acknowledges partial support
from Fondo de Apoyo a la Investigaci\'on (Universidad Aut\'onoma de San
Luis Potos{\'\i}) through Grant No. C00-FAI-03-8-18.

\appendix

\section{}

In this Appendix we give for completeness the amplitudes for the RC of the
decay~(\ref{eq:e1}). All of them are also given in Ref.~\cite{Flores}. The
uncorrected transition amplitude $M_{0}$ for process~(\ref{eq:e1}) is
\begin{equation} 
M_{0}=\frac{G_V}{\sqrt{2}}\left[ \overline{u}_{B}\left(
p_{2}\right) W_{\mu}\left( p_{1},p_{2}\right) u_{A}\left( p_{1}\right)
\right] \left[ \overline{u}_\ell (l) O_\mu v_\nu \left( p_\nu \right)
\right] , \label{eq:eA1} 
\end{equation} 
where 
\begin{eqnarray} 
W_{\mu}(p_1,p_2) &=& f_1(q^2) \gamma_\mu + \frac{f_2(q^2)}{M_1}
\sigma_{\mu\nu} q_\nu + \frac{f_3(q^2)}{M_1} q_\mu \nonumber \\
&  & \mbox{} + \left[g_1(q^2) \gamma_\mu + \frac{g_2(q^2)}{M_1}
\sigma_{\mu\nu} q_\nu + \frac{g_3(q^2)}{M_1} q_\mu \right] \gamma_5 .
\label{eq:eA2}
\end{eqnarray}
Here $O_{\mu}=\gamma_{\mu}( 1+\gamma_{5})$ and $q$ is the four-momentum
transfer.

The model-independent part of the virtual radiative corrections has the
amplitude 
\begin{equation}
M_v=\frac{\alpha}{2\pi}\left[ M_{0}\phi \left( E\right) +M_{p_{1}} 
\phi^\prime \left( E\right) \right] , \label{eq:eA4}
\end{equation}
where $\phi \left(E\right)$ and $\phi^\prime \left( E\right)$ were
given in Eqs.~(\ref{eq:e11}) and (\ref{eq:e12}), respectively. $M_{p_1}$
is
\begin{equation}
M_{p_{1}}=\left( \frac{E}{mM_{1}}\right) \frac{G_V}{\sqrt{2}}\left[ 
\overline{u}_{B}W_{\lambda} u_{A}\right] \left[ \overline{u}_{\ell} 
{\not{\!}p}_{1} O_{\lambda}v_{\nu}\right] . \label{eq:eA5}
\end{equation}

The model-dependent part of the virtual radiative corrections is absorbed
into $M_{0}$ through the definition of effective form factors $f_1^\prime$
and $g_1^\prime$. This fact is denoted by putting a prime on $M_{0}$. The
bremsstrahlung transition amplitude $M_{B}$ is obtained following the Low
theorem~\cite{Low},
\begin{eqnarray}
M_{B} &=&\frac{eG_V}{\sqrt{2}}\left[ \overline{u}_{B}W_{\lambda
} u_{A}\right] \left[ \overline{u}_{l}O_{\lambda}v_{\nu}\right] \left[
\frac{l \cdot \epsilon}{l \cdot k}-\frac{p_{1}\cdot \epsilon}{ p_{1}\cdot
k} \right] + \frac{eG_V}{\sqrt{2}}\left[ \overline{u}_{B}W_{\lambda}u_{A}
\right] \left[ \frac{\overline{u}_{\ell}\not{\epsilon}\not
{k}O_{\lambda}v_{\nu}}{ 2 l \cdot k}\right]  \nonumber \\
&\equiv &M_{a}+M_{b} . \label{eq:eA6}
\end{eqnarray}
Within our approximation the Low theorem guarantees that no 
model-dependence appears here.

\section{}

In order to make this paper self-contained, we reproduce here all the
coefficients which appear in our final results. For Eq.~(\ref{eq:e7}) they
come from Ref.~\cite{Flores},
\begin{eqnarray}
A_{0}^\prime &=&Q_{1}EE_{\nu}^{0}-Q_{2}Ep_{2}\left( p_2 + l
y_{0}\right) -Q_{3} l \left( p_{2}y_0 + l \right) + Q_{4}E_{\nu}^{0}p_{2} l
y_0-Q_5 p_2^2 l y_0 \left( p_2 + l y_{0}\right) , \label{eq:eB1} \\
&&  \nonumber \\
A_{1}^\prime &=&D_{1}EE_{\nu}^{0}-D_{2} l \left( p_2 y_0 + l
\right) , \label{eq:eB2} \\
&&  \nonumber \\
A_{1}^{\prime \prime} &=&D_{1}EE_{\nu}^{0} . \label{eq:eB3}
\end{eqnarray}

The coefficients $Q_i$, $i=1,\ldots,7$ are given in Eqs.~(B1)-(B7) of
Ref.~ \cite{Flores}, respectively. For completeness we reproduce them
here.

\begin{eqnarray}
Q_{1} &=&F_{1}^{2}\left[ \frac{2E_{2}-M_{2}}{M_{1}}\right] +\frac{1}{2}
F_{2}^{2}\left[ \frac{M_{2}+E_{2}}{M_{1}}\right] +F_{1}F_{2}\left[ \frac{
M_{2}+E_{2}}{M_{1}}\right] +F_{1}F_{3}\left[ 1+\frac{M_{2}}{M_{1}}\right] 
\nonumber \\
&&\times \left[ 1-\frac{E_{2}}{M_{1}}\right] +F_{2}F_{3}\left[ \frac{
M_{2}+E_{2}}{M_{1}}\right] \left[ 1-\frac{E_{2}}{M_{1}}\right]
+G_{1}^{2}\left[ \frac{2E_{2}+M_{2}}{M_{1}}\right]  \nonumber \\
&& \mbox{} -\frac{1}{2}G_{2}^{2}\left[ \frac{M_{2}-E_{2}}{M_{1}}\right]
+G_{1}G_{2}\left[ \frac{M_{2}-E_{2}}{M_{1}}\right] +G_{1}G_{3}\left[ \frac{
M_{2}}{M_{1}}-1\right] \left[ 1-\frac{E_{2}}{M_{1}}\right]  \nonumber \\
&& \mbox{} -G_{2}G_{3}\left[ \frac{M_{2}-E_{2}}{M_{1}}\right] \left[
1-\frac{E_{2}}{M_1} \right] +M_{1}^{2}Q_{5}\left\{ \left[
\frac{M_{1}-E_{2}}{M_{1}}\right]^2 - \frac{1}{2}
\frac{q^2}{M_1^2}\right\} , \label{eq:eB4} \\
&& \nonumber \\
Q_2 &=&-\frac{F_1^2}{M_1}-\frac{G_{1}^{2}}{M_{1}}-\frac{F_{1}F_{2}}{
M_1}+\frac{G_1G_2}{M_1}+\frac{F_{1}F_{3}}{M_{1}}\left[ 1+\frac{M_{2} 
}{M_{1}}\right] +\frac{F_{2}F_{3}}{M_{1}}\left[ \frac{M_{2}+E_{2}}{M_{1}}
\right]  \nonumber \\
&& \mbox{} +\frac{G_{1}G_{3}}{M_{1}}\left[ \frac{M_{2}}{M_{1}}-1\right]
-\frac{G_{2}G_{3}}{M_{1}}\left[ \frac{M_{2}-E_{2}}{M_{1}}\right]
+2\frac{F_{1}G_{1}}{M_{1}} + M_{1}Q_{5}\left[
\frac{M_{1}-E_{2}}{M_{1}}\right] , \label{eq:eB5} \\
&& \nonumber \\
Q_{3} &=&Q_{1}-2F_{1}^{2}\left[ \frac{E_{2}-M_{2}}{M_{1}}\right]
-2G_{1}^{2}\left[ \frac{E_{2}+M_{2}}{M_{1}}\right] -M_{1}^{2}Q_{5}\left\{
\left[1-\frac{E_{2}}{M_1} \right]^2 - \frac{q^2}{M_1^2} \right\} ,  
\label{eq:eB6} \\ && \nonumber \\
Q_{4} &=&Q_{2}-4\frac{F_{1}G_{1}}{M_{1}} , \label{eq:eB7} \\
&& \nonumber \\
Q_{5} &=&\frac{F_{3}^{2}}{M_{1}^{2}}\left[ \frac{M_{2}+E_{2}}{M_{1}} \right]
- \frac{G_{3}^{2}}{M_{1}^{2}}\left[ \frac{M_{2}-E_{2}}{M_{1}}\right] -2\frac{
F_{1}F_{3}}{M_{1}^{2}}+2\frac{G_{1}G_{3}}{M_{1}^{2}} , \label{eq:eB8} \\
&& \nonumber \\
Q_{6} &=&F_{1}^{2}\left[ \frac{E_{2}-M_{2}}{M_{1}}-\frac{p_{2}\beta y_{0}}{
M_{1}}\right] +G_{1}^{2}\left[ \frac{E_{2}+M_{2}}{M_{1}}-\frac{p_{2}\beta
y_{0}}{M_{1}}\right] +2F_{1}G_{1}\left[ \frac{E_{2}-p_{2}\beta y_{0}}{M_{1}}
\right]  \nonumber \\
&& \mbox{} +\left( G_{1}G_{2}-F_{1}F_{2}\right) \left[ \frac{p_{2}\beta
y_{0}}{M_{1}} \right] +F_{2}G_{2}\left[ -1+\left(
1+\beta^{2}\right) \frac{E}{M_{1}} + \frac{E_{2}}{M_{1}}+\frac{p_{2}\beta
y_{0}}{M_{1}}\right]  \nonumber \\
&& \mbox{} +F_{1}G_{2}\left[ -1+\frac{M_{2}}{M_{1}}+\left(
1+\beta^{2}\right) \frac{E}{M_{1}} + \frac{p_{2}\beta y_{0}}{M_{1}}\right]
-G_{1}F_{2}\left[ -1-\frac{M_{2}}{M_{1}}+\right.  \nonumber \\
&&\left. \left( 1+\beta^{2}\right) \frac{E}{M_{1}}+\frac{p_{2}\beta y_{0}}{
M_{1}}\right] -F_{3}G_{3}\left[ \frac{m^{2}}{M_{1}^{2}}\left( 1-\frac{E_{2}}{
M_{1}}-\left( 1-\beta^{2}\right) \frac{E}{M_{1}}+\frac{p_{2}\beta y_{0}}{
M_{1}}\right) \right]  \nonumber \\
&& \mbox{}+F_{1}G_{3}\left[ \frac{m^{2}}{M_{1}E}\left(
-1+\frac{M_{2}}{M_{1}} + \frac{E}{M_{1}} \right) \right] -F_{3}G_{1}\left[
\frac{m^{2}}{M_{1}E}\left( -1- \frac{M_{2}}{M_{1}} + \frac{E}{M_{1}}
\right) \right]  \nonumber \\
&& \mbox{} -\left( F_{2}G_{3}+F_{3}G_{2}\right) \left[
\frac{m^{2}}{M_{1}E}\left(\frac{M_{1}-E_{2}-E}{M_{1}}\right) \right] ,
\label{eq:eB9} \\
&&  \nonumber \\
Q_{7} &=&F_{1}^{2}\left[ \frac{\left( M_{1}+M_{2}\right) \left(
E_{2}-M_{2}\right)}{M_{1}E}\right] +G_{1}^{2}\left[ \frac{\left(
M_{1}-M_{2}\right) \left( E_{2}+M_{2}\right)}{M_{1}E}\right]  \nonumber \\
&& \mbox{} + 2F_{1}G_{1}\left[ \frac{M_{1}\left(-M_1 + E_2 + 
2E\right) -m^{2}}{M_{1}E} \right] +F_{1}G_{2}\left(
\frac{E_{2}-M_{2}}{M_{1}}\right) \nonumber \\
&& \times \left( \frac{M_{1}-2E-E_{2}}{E}\right) -G_{1}F_{2}\left( \frac{
E_{2}+M_2}{M_1}\right) \left( \frac{M_1-2E-E_2}{E}\right) \nonumber \\
&& \mbox{} +F_{3}G_{1}\left( \frac{E_{2}+M_{2}}{M_{1}}\right) \left(
\frac{m^{2}}{M_{1}E}\right) -G_{3}F_{1}\left( \frac{E_{2} -
M_{2}}{M_{1}}\right) \left(\frac{m^{2}}{M_{1}E}\right)  \nonumber \\
&& \mbox{} +\left( F_{1}F_{2}-G_{1}G_{2}\right) \left(
\frac{E_{2}^{2}-M_{2}^{2}}{M_{1}E}\right) . \label{eq:eB10}
\end{eqnarray}

The coefficients $D_j$, $j=1,\ldots,4$ are 
\begin{eqnarray}
D_1 &=&{f_1^\prime}^2+ 3 {g_1^\prime}^2 , \label{eq:eB11} \\
&&  \nonumber \\
D_2 &=&{f_1^\prime}^2-{g_1^\prime}^2 , \label{eq:eB12} \\
&&  \nonumber \\
D_3 &=&2 (f_1^\prime g_1^\prime - {g_1^\prime}^2) , \label{eq:eB13}
\\
&&  \nonumber \\
D_4 &=&2 (f_1^\prime g_1^\prime + {g_1^\prime}^2) . \label{eq:eB14}
\end{eqnarray}
$E_{\nu}^{0}$ and $y_{0}$ were defined in Eqs.~(\ref{eq:e5}) and
(\ref{eq:e6}), respectively.

The functions $\theta_i$ which appear in 
Eqs.~(\ref{eq:e48})-(\ref{eq:e51}) corresponding to the TBR of the DP are
given by 
\begin{equation}
\theta_i=\frac{1}{p_2} \left( T_i^++T_i^- \right) , \label{eq:eB15}
\end{equation}
where $i=2,\ldots ,16$, and 
\begin{eqnarray}
T_{2}^{\pm} &=&\pm \frac{1\mp a^\pm}{(1 \pm \beta)(1 + \beta a^\pm)} \ln
\left[ \frac{1\mp \beta}{1-\beta
x_{0}}\right] \pm \frac{(1 \pm x_0) \ln (1 \pm x_0)}{(1 \pm \beta)(1 -
\beta x_0)} \nonumber \\
&&\mbox{} \pm \frac{1\pm a^\pm}{(1 \mp \beta)(1 + \beta a^\pm)}
\ln (1 \pm a^\pm) -\frac{(x_0 + a^\pm) \ln (\pm x_0 \pm a^\pm)}{(1 + \beta
a^\pm)(1 - \beta x_0)} , \label{eq:eB16} \\
&& \nonumber \\
T_{3}^{+} &=&T_{3}^{-}=\frac{1}{2\beta} \left\{ L\left[ \frac{1-\beta}{
1-\beta x_{0}}\right] -L\left[ \frac{1-\beta x_{0}}{1+\beta} \right]
-L\left[ \frac{1+\beta a^{-}}{1-\beta x_{0}}\right] +L\left[ \frac{1+\beta
a^{-}}{1+\beta} \right] \right. \nonumber \\
&& \mbox{} + \left. L\left[ \frac{1-\beta x_{0}}{1+\beta a^{+}}\right]
-L\left[ \frac{ 1-\beta} {1+\beta a^{+}}\right] +\ln \left[ \frac{1-\beta
x_{0}}{1-\beta} \right] \ln \left[ \frac{1+\beta a^{+}}{1+\beta} \right]
\right\} , \label{eq:eB17} \\
&& \nonumber \\
T_4^\pm &=& (x_0 \pm 1) \ln (1 \pm x_0) \pm (1 \pm a^\pm) \ln (1 \pm
a^\pm) - (x_0 + a^\pm) \ln (\pm x_0 \pm a^\pm) , \label{eq:eB18} \\
&& \nonumber \\
T_{5}^\pm &=&-\frac{1}{2} \left\{ (1 - x_0^2) \ln (1 \pm x_0) + (x_0 \mp
1) a^\pm + 1 - (1-a^{\pm 2}) \ln (1 \pm a^\pm) \right.   \nonumber \\
&& \mbox{} + \left. (x_0^2 - a^{\pm 2}) \ln [\pm (x_0 + a^\pm)] \right\} ,
\label{eq:eB19} \\
&& \nonumber \\
T_{6}^\mp &=&\left[-l + p_2 \pm \frac{\beta E_\nu^0 (x_0+a^\mp)}{1 + \beta
a^\mp} \right] I_4 \pm \frac{\beta E_\nu^0 (x_0 + a^\mp)}{(1 + \beta a^\mp)^2}
I_{1}+\left[ E_\nu^0 - \frac{\beta E_\nu^{0} (x_0 + a^\mp)}{1 + \beta 
a^\mp} \right] J_{4}  \nonumber \\
&& \mbox{}-\frac{\beta E_\nu^0 (x_0 + a^\mp)}{(1 + \beta a^\mp)^2} 
J_1 \pm \frac{E_\nu^0 (x_0 + a^\mp)}{(1 + \beta a^\mp)^2} I_2^\mp -
\frac{E_\nu^0 (x_0 + a^\mp)}{(1 + \beta a^\mp)^2} J_2^\mp , 
\label{eq:eB20} \\
&& \nonumber \\
T_7^\pm &=&\left[ p_{2}-l \mp \frac{\beta E_\nu^0 (x_0 + a^\pm)}{1 + \beta
a^\pm}\right] I_1 \mp \frac{E_\nu^0 (x_0 + a^\pm)}{1 + \beta a^\pm}
I_2^\pm + \left[E_\nu^0 - \frac{\beta E_\nu^0 (x_0 + a^\pm)}{1 + \beta
a^\pm} \right] J_1 \nonumber \\ && \mbox{} -\frac{E_\nu^0 \left(
x_{0}+a^\pm \right)}{1+\beta a^{\pm }} J_2^\pm , \label{eq:eB21} \\
&& \nonumber \\
T_8^\pm &=&-2 (l -p_2+E_\nu^0 x_0) \mp E_\nu^0 (x_0 + a^\pm) I_2^\pm 
- E_\nu^0 (x_0 + a^\pm) J_2^\pm , \label{eq:eB22} \\
&& \nonumber \\
\frac{T_9^\pm}{4 l^2} &=&-\frac{3E}{2l^2} (l -p_2+E_\nu^0 x_0) +\left[
\frac{3(l -p_2)}{4\beta l} + \frac{3E_\nu^0 p_2}{4l^2}+\beta G^\pm \right]
I_1 \mp \frac{(E_\nu^0)^2 ( x_{0}+a^{\pm})^{2}}{4l^{2} (1+\beta a^\pm)}
I_3^\pm \nonumber \\
&& \mbox{} -\frac{(E_\nu^0)^2 (x_{0}+a^\pm)^{2}}{4l^2 (1 + \beta a^\pm)}
J_3^\pm + G^\pm I_2^\pm + \left[- \frac{3E_\nu^0}{4\beta l}+\frac{3E_\nu^0
(E_\nu^0+ l x_0)}{4l^2}\pm \beta G^{\pm }\right] J_{1}\pm G^\pm J_2^\pm ,
\label{eq:eB23} \\
&&  \nonumber \\
T_{10}^{\mp} &=& \frac{1}{3} (x_0^3 \mp 1) \ln(1 \mp x_0) +
\frac{1}{3} \left[ (a^\mp)^3 \mp 1 \right] \ln(1 \mp a^{\mp}) - 
\frac{1}{3}\left[ x_{0}^{3}+(a^\mp)^3 \right]
\ln \left[ \mp (x_{0}+a^{\mp}) \right] \nonumber \\
&& \mbox{} + \frac{1}{6} (1-x_0^{2})( a^{\mp} \pm 1) -\frac{1}{3} (
x_{0}\pm 1) \left[ 1-  (a^\mp)^2 \right] , \label{eq:eB24} \\
&&  \nonumber \\
T_{11}^{+} &=&T_{11}^{-}=\frac{1}{2p_{2}\beta }\left\{ E_{\nu }^{0}\left[
( 1-\beta x_{0}) J_{4}-J_{1}\right] -( \beta E_{\nu
}^{0}+l -p_{2}) I_{4}+(l -p_{2}) I_{1}\right\} ,
\label{eq:eB25} \\
&&  \nonumber \\
T_{12}^{+} &=&T_{12}^{-}=\frac{1}{2p_{2}\beta }\left[ E_{\nu }^{0}(1 
- \beta x_{0}) J_{1}+2E_{\nu }^{0}x_{0}+2(l -p_{2})
- (\beta E_\nu^0+l -p_2) I_{1}\right] , \label{eq:eB26} \\
&&  \nonumber \\
T_{13}^{+} &=&T_{13}^{-}=-\frac{1}{2p_{2}}E_{\nu }^{0} (1-x_{0}^{2}) ,
\label{eq:eB27} \\
&&  \nonumber \\
T_{14}^{\pm } &=&E_{\nu }^{0}\left[ 1+x_{0}^{2}+2a^\pm (x_{0}\mp 1) \pm
a^{\pm } ( x_{0}+a^{\pm })(I_{2}^{\pm }\pm J_2^\pm) \right] ,
\label{eq:eB28} \\
&&  \nonumber \\
T_{15}^{\pm } &=&3E_{\nu }^{0}\left[ 2p_{2}( 1+y_{0}) +l
( 1-x_{0}^{2}) \right] - (E_\nu^0)^2 (x_0+a^\pm)^{2} (J_3^\pm \pm
I_{3}^\pm) \nonumber \\
&& \mbox{} -2l E_{\nu }^{0}( x_{0}+a^{\pm }) a^\pm (J_{2}^{\pm} \pm 
I_{2}^\pm) , \label{eq:eB29} \\
&&  \nonumber \\
T_{16}^{\pm } &=&4l^{2}\left[ \frac{3}{2\beta^{2}}\left[ 2( l
-p_{2}+E_{\nu }^{0}x_{0}) +\beta E_{\nu }^{0} (1-x_{0}^{2})
\right] + \left( -\frac{3 ( l -p_{2}+\beta E_{\nu }^{0})}{
2\beta^{2}}\right. \right. \nonumber \\
&& \mbox{} -\left. p_2 ( 1+y_0) +\frac{p_2(
E_\nu^0)^2}{2l^2}\right) I_{1}-\frac{( E_{\nu}^0)^2 (x_0 + a^{\pm
})^{2}}{2l ( 1+\beta a^{\pm }) }(\beta J_1+J_2^\pm \pm \beta I_1\pm 
I_2^\pm) \nonumber \\
&& \mbox{} +\left. \left( \frac{3E_{\nu }^{0} ( 1-\beta x_0)}{2
\beta^{2}}+ \frac{( E_{\nu }^{0})^{2}(E_\nu^0+l x_{0})}{2l^2}\right)  
J_{1}\right] . \label{eq:eB30}
\end{eqnarray}

The following definitions are used in the above expressions
\begin{eqnarray}
x_{0} &=&-\frac{p_{2}y_{0}+l }{E_{\nu }^{0}} , \qquad a^{\pm }=\frac{
E_{\nu }^{0}\pm p_{2}}{l} , \label{eq:eb31a} \\
&&  \nonumber \\
I_{1} &=&\frac{2}{\beta }\text{arctanh}\beta , \quad \quad I_{2}^{\pm
}=\ln \left| \frac{a^{\pm }+1}{a^{\pm }-1}\right| , \label{eq:eb31b} \\
&&  \nonumber \\
I_{3}^{\pm } &=&\frac{2}{a^{\pm 2}-1} , \quad \quad \quad I_{4}=\frac{2}{
1-\beta^{2}} , \label{eq:eb31c} \\
&&  \nonumber \\
J_1 &=&-\frac{1}{\beta}\left\{ \ln \left[ \frac{1+\beta}{1-\beta x_0}
\right] +\ln \left[ \frac{1-\beta }{1-\beta x_{0}}\right] \right\} ,
\label{eq:eb31d} \\
&&  \nonumber \\
J_{2}^{\pm } &=&\ln \left| \frac{a^{\pm }-1}{a^{\pm }+x_{0}}\right| +\ln
\left| \frac{a^{\pm }+1}{a^{\pm }+x_{0}}\right| , \label{eq:eb31e} \\
&&  \nonumber \\
J_{3}^{\pm } &=&-2\left[ \frac{a^{\pm }}{a^{\pm 2}-1}-\frac{1}{a^{\pm }+x_{0}
}\right] , \label{eq:eb31f} \\
&&  \nonumber \\
J_{4} &=&\frac{2}{\beta }\left[ \frac{1}{1-\beta^{2}}-\frac{1}{1-\beta x_{0}
}\right] , \label{eq:eb31g} \\
&&  \nonumber \\
G^{\pm } &=&\mp \frac{\beta (E_\nu^0)^2 \left( x_{0}+a^{\pm}\right)^{2}
}{4l^2 \left( 1+\beta a^{\pm }\right)^{2}}\mp \frac{a^{\pm }\left(
a^{\pm 2}-1\right) }{4\left( 1+\beta a^{\pm }\right) } . \label{eq:eb31h}
\end{eqnarray}

The sums $\theta^\prime +\theta^{\prime \prime \prime}$ and
$\theta^{\prime \prime}+\theta^{\rm IV}$ which appear in Eq.~(\ref
{eq:e54}) are
\begin{eqnarray}
\theta^\prime +\theta^{\prime \prime \prime} &=&\frac{p_2 l}{
2}\left[ -E_{\nu }^{0}\left( 1-\beta^{2}\right) \theta_2+\left( E_{\nu
}^{0}-\frac{1+\beta^{2}}{2}E\right) \theta_3+\frac{E}{2}\theta
_{4}\right.  \nonumber \\
&& \mbox{} + \left. \frac{l}{2}\theta_5+\frac{1-\beta^{2}}{2}\theta_6 
-\frac{2E- E_\nu^0}{2E} \theta_7 + \frac{1}{2}\theta_8-\frac{1}{4E}\theta
_{9}\right] , \label{eq:eB38}
\end{eqnarray}

\begin{equation}
\theta^{\prime \prime} +\theta^{\rm IV} = \frac{p_2 l}{2}\left[
\theta_{0}-\left( E+E_{\nu}^{0}+\beta p_{2}y_{0}\right) \theta_{3}+\left(
E_{\nu}^{0}+E\right) \theta_{4}+l \theta_{5}\right] . \label{eq:eB39}
\end{equation}

The explicit form of the photon integrals corresponding to the FBR of the DP
is 
\begin{eqnarray}
\theta_{2F} &=&\frac{1}{\beta p_{2}}\left[ \frac{I_{2}^{\,\,-}}{b^{-}}-
\frac{I_{2}^{\,\,+}}{b^{+}}+\frac{E^{2}}{m^{2}}\left(
I_{2}^{\,\,+}-I_{2}^{\,\,-}+\beta \ln \left| \frac{I_{3}^{\,\,-}}{
I_{3}^{\,\,+}}\right| \right) \right] +\frac{2I_{1}}{Eb^{-}b^{+}} ,
\label{eq:eB40} \\
&&  \nonumber \\
\theta_{3F} &=&\frac{I_{1}}{p_{2}}\ln \left| \frac{b^{+}}{b^{-}}\right| +
\frac{1}{\beta p_{2}}\left[ L\left( \frac{1-\beta }{b^{-}}\right) -L\left( 
\frac{1-\beta }{b^{+}}\right) + L\left( \frac{1+\beta
}{b^{+}}\right) -L\left( \frac{1+\beta }{b^{-}}\right) \right] ,
\label{eq:eB41} \\
&&  \nonumber \\
\theta_{4F} &=&\frac{1}{p_{2}}\left[a^{+} I_{2}^{\,\,+} -
a^{-}I_{2}^{\,\,-} + \ln \left| \frac{I_{3}^{\,\,-}}{
I_{3}^{\,\,+}}\right| \right] , \label{eq:eB42} \\
&&  \nonumber \\
\theta_{5F} &=&\frac{1}{2p_{2}}\left[(1-a^{+\,\,2}) I_{2}^{\,\,+}- 
(1-a^{-\,\,2}) I_{2}^{\,\,-}+\frac{4p_{2}}{l} \right] , \label{eq:eB43} \\
&&  \nonumber \\
\theta_{6F} &=&2\frac{y_{0}^{-}}{( b^{-})^{2}}( I_{2}^{\,\,-}+\beta
I_{1}) -2\frac{y_{0}^{+}}{( b^{+})^{2}} ( I_{2}^{\,\,+}+\beta
I_{1}) + 2\left[ 2 + \beta \left( \frac{y_0^-}{b^-} - \frac{y_0^+}{b^+}
\right) \right] I_{4} , \label{eq:eB44} \\
&&  \nonumber \\
\theta_{7F} &=&2\left[ 2I_{1}+\frac{y_{0}^{-}}{b^{-}} (\beta 
I_{1}+I_{2}^{\,\,-}) -\frac{y_{0}^{+}}{b^{+}} (\beta 
I_{1}+I_{2}^{\,\,+}) \right] , \label{eq:eB45} \\
&&  \nonumber \\
\theta_{8F} &=&2\left[4 +(y_{0}^{-})I_{2}^{\,\,-} - (y_{0}^{+})
I_{2}^{\,\,+} \right] , \label{eq:eB46} \\
&&  \nonumber \\
\theta_{9F} &=&24E+2\left[ 6 ( E_{\nu }^{0}-E) +\beta 
(G_{F}^{-}+G_{F}^{+}) \right] I_{1}+2(G_{F}^{-} I_{2}^{-} +
G_{F}^{+}I_{2}^{+})  \nonumber \\
&& \mbox{} +2p_{2}\left[ \frac{(y_0^-)^{2}}{b^{-}} I_3^{\,\,-}
- \frac{(y_{0}^{+})^{2}}{b^{+}}I_{3}^{\,\,+}\right] , \label{eq:eB47} \\
&&  \nonumber \\
\theta_{10F} &=&\frac{1}{3p_{2}}\left[ 2( a^{-2}-a^{+2})
-a^{-3}I_{2}^{\,\,-}+a^{+3}I_{2}^{\,\,+}+\ln \left| \frac{I_{3}^{\,\,-}}{
I_{3}^{\,\,+}}\right| \right] , \label{eq:eB48} \\
&&  \nonumber \\
\theta_{11F} &=&\frac{2( I_4-I_{1}) }{\beta p_{2}} , \label{eq:eB49} \\
&&  \nonumber \\
\theta_{12F} &=&\frac{2(I_1-2)}{\beta p_2} , \label{eq:eB50} \\
&&  \nonumber \\
\theta_{13F} &=&0 , \label{eq:eB51} \\
&&  \nonumber \\
\theta_{14F} &=&2\left[( 2-a^{-}I_{2}^{\,\,-})(y_0^{-}) - (2-a^{+}
I_{2}^{\,\,+}) (y_{0}^{+}) \right] , \label{eq:eB52} \\
&&  \nonumber \\
\theta_{15F} &=&24E_{\nu }^{0}+4 l \left[a^{-} y_{0}^{-} I_{2}^{\,\,-} -
a^{+} y_{0}^{+} I_{2}^{\,\,+}\right] + 2p_{2}\left[(y_{0}^{-})^2
I_{3}^{\,\,-}- (y_{0}^{+})^{2} I_{3}^{\,\,+}\right] , \label{eq:eB53} \\
&&  \nonumber \\
\theta_{16F} &=&24E^{2}( I_{1}-2) +8 [(E_\nu^0)^2 - 
2E^{2} \beta^{2}] I_{1}  \nonumber \\
&& \mbox{}+4 l p_{2}\left[ \frac{( y_{0}^{-})^{2}}{b^{-}} (\beta 
I_{1}+I_{2}^{\,\,-}) -\frac{(y_{0}^{+})^2}{b^+}
(\beta I_{1}+I_{2}^{\,\,+}) \right] , \label{eq:eB54}
\end{eqnarray}
where $a^{\pm }$, $I_1$, $I_2^{\pm}$, $I_3^{\pm}$, $I_4$ are given in
Eqs.~(\ref{eq:eb31a})-(\ref{eq:eb31c}) and
\begin{eqnarray}
b^\pm = 1+\beta a^{\pm } , \quad \quad y_{0}^{\pm }=y_{0}\pm a^{\pm } ,
\nonumber
\end{eqnarray}
\begin{eqnarray}
G_{F}^{\,\pm }=\mp \beta \left( 2Ea^{\pm }+p_{2}\frac{y_{0}^{\pm
}}{b^{\pm }} \right) \frac{y_{0}^{\pm }}{b^{\pm }} . \nonumber
\end{eqnarray}

\section{}

In this appendix we give a brief review of the procedure of Ref.~\cite
{Ginsberg} to extract the infrared divergence and the finite terms that
come along with it. This procedure can be adapted to our case by
expressing the differential decay rate of the bremsstrahlung radiative
corrections $d\Gamma_B$ in terms of the invariant mass $\eta
=(p_{\nu}+k)^{2}$ as follows
\begin{eqnarray}
d\Gamma_{B} &=&\frac{M_{2}mm_{\nu}}{\left( 2\pi \right)^{8}}\frac{
dE\;dE_{2}\;d\Omega_{\ell}\;d\phi_{2}}{4}\int_{\eta_{\min}}^{\eta
_{\max}}d\eta \frac{d^{3}k}{\omega}\frac{d^{3}p_{\nu}}{E_{\nu}} 
\sum_{{\rm spins}}\left| M_{B}\right|^{2}\delta^{4}\left(
p_{1}-p_2-l-p_{\nu}-k\right) . \label{eq:eC1}
\end{eqnarray}

For the TBR and in the CM-frame of the decaying particle, $\eta =2p_2 l
\left( y_{0}-y\right) $, and $\eta_{\max}=2p_2 l \left( y_{0}+1\right) $
and $\eta_{\min}=\lambda^2$, with $\lambda \rightarrow 0$. Here $\lambda$
is a small mass assigned to the photon. From Eq.~(\ref{eq:eA6}) we have
$M_{B}=M_{a}+M_{b}$, and (we do not consider the polarization here) the
divergent part of $\sum_{{\rm spins}}\left| M_{B}\right|^{2}$ is contained
in the first two terms of the square brackets of
\begin{eqnarray}
\sum_{{\rm spins}}\left| M_{a}\right|^{2} &=&\frac{e^{2}G_V^{2}}{2}\frac{
2M_{1} }{M_{2}mm_{\nu}}\sum_{\epsilon}\left( \frac{l \cdot \epsilon}{l
\cdot k}-\frac{p_{1}\cdot \epsilon}{p_{1}\cdot k}\right)^{2}  \nonumber \\
&& \times \left[ D_{1}EE_{\nu}^{0}-D_{2} {\bf l}\cdot \left({\bf p}_2 +
{\bf l} \right) -D_{1}E\omega -D_{2} {\bf l}\cdot {\bf k}\right].
\label{eq:eC2}
\end{eqnarray}

Because ${\bf l}\cdot \left( {\bf p}_2 + {\bf l}\right) =l p_2
y_0+l^2-\eta /2$ (choosing the $z$-axis along ${\bf l}$), we can split
$d\Gamma_B$ as
\begin{eqnarray}
d\Gamma_{B} &=&\frac{1}{2}\frac{G_V^{2}}{2} \frac{dE \; dE_{2}
\; d\Omega_\ell \; d\phi_2} {\left( 2\pi \right)^{8}} 4 M_1
\frac{\alpha}{\pi} 
\left\{ A_{1}^\prime \frac{1}{4} \lim_{\lambda \rightarrow 0} 
\int_{\lambda^{2}}^{\eta_{\max}}d\eta \frac{1}{2\pi}\frac{d^{3}k}{
\omega} \frac{d^{3}p_{\nu}}{E_{\nu}} \delta^{4}\left( p_{1}-p_{2}-l
-p_{\nu}-k\right)  \right. \nonumber \\
&  & \mbox{} \times
\left[ \frac{2l \cdot p_{1}}{(p_{1}\cdot k)(l \cdot k)} -
\frac{m^{2}}{\left(l \cdot k\right)^{2}}-\frac{M_{1}^2}{w \left( p_{1}
\cdot k \right)^2} \right] \nonumber \\
&  & \mbox{} + \frac{1}{4}\int_{0}^{\lambda_{\max}}d\eta \frac{1}{2\pi}
\frac{d^{3}k}{\omega} \frac{d^{3}p_{\nu}}{E_{\nu}}\delta^{4}\left(
p_{1}-p_{2}-l -p_{\nu}-k\right) \sum_{\epsilon}\left( \frac{l \cdot
\epsilon}{l \cdot k} - \frac{p_1 \cdot \epsilon}{p_1 \cdot k}\right)^{2} 
\nonumber \\
&&\times \left[- D_{1}E\omega -D_{2} \left( {\bf l} \cdot {\bf k} +
\frac12 \eta \right) \right] \left. \sum_{\epsilon ,s}\left[ \left|
M_{b}\right|^{2}+2 \mathop{\rm Re} \left[ M_{a}\right] \left[
M_{b}\right]^{\dagger}\right]
\right\} , \label{eq:eC3}
\end{eqnarray}
where $A_{1}^\prime$ is given in Eq.~(\ref{eq:eB2}). In the first
integral we perform the sum over polarizations indicated in
Eq.~(\ref{eq:eC2}) in covariant form. This allow us to identify this
integral with the divergent integral $I_{0}$ of Ref.~\cite{Ginsberg},
namely,
\begin{equation}
I_{0}\left( E,E_{2}\right) = \frac14 \lim_{\lambda \rightarrow 0} 
\int_{\lambda^2}^{\eta_{\max}}d\eta \left[ 2(l \cdot p_{1})I_{11}\left( l
,p_{1}\right) -m^{2}I_{20}\left( l ,p_1\right) -M_{1}^{2}I_{02}\left(
l ,p_1 \right) \right] , \label{eq:eC4}
\end{equation}
with 
\begin{equation}
I_{m\,n}\left( l, p_1 \right) =\frac{1}{2\pi}\int \frac{d^{3}p_{\nu}}{
E_{\nu}}\frac{d^{3}k}{\omega}\frac{\delta^{4}\left( p_{1}-p_{2}-l
-p_{\nu}-k\right)}{\left( l \cdot k\right)^{m}\left( p_{1}\cdot
k\right)^{n}} , \label{eq:eC5}
\end{equation}
which are invariant integrals. The final form of $I_{0}$ is 
\begin{eqnarray}
I_{0}\left( E,E_{2}\right) &=&\frac{1}{\beta}\text{arctanh} \beta \left[
2\ln \left( \frac{2 l}{\lambda}\right) +\ln \left( \frac{m\,\eta_{m}^{2}}{
4\left( E+l \right) r_{+}}\right) \right] -\frac{1}{\beta}L\left(-
\frac{a^{2}}{4r_{+}}\right) \nonumber \\
&& \mbox{} + \frac{1}{\beta} L\left( -\frac{4r_{-}}{a^{2}}\right) -2\ln
\left( \frac{m}{\lambda}\right) -\ln \left( \frac{\eta_m^2}{2mE_\nu^0
\left( q^{2}-m^{2}\right)} \right) , \label{eq:eC6}
\end{eqnarray}
where 
\begin{eqnarray}
\left( E+l \right) r_{\pm} & = &\left[ E_{\nu}^{0} l^2 \left(
q^{2}-m^{2}\right) -a^{2}E/4\right] \nonumber \\
& & \mbox{} \pm \left\{ \left[ E_{\nu}^{0} l^2 \left( q^{2}-m^{2}\right)
-a^{2}E/4\right]^{2}-m^{2}a^{4}/16\right\}^{1/2} , \label{eq:eC7}
\end{eqnarray}
\begin{equation}
a^{2} = \eta_{m}\left( 4p_{2} l -\eta_{m}\right) , \label{eq:eC8}
\end{equation}
and
\begin{equation}
q^{2} = M_{1}^2-2M_{1}E_{2}+M_{2}^2 . \label{eq:eC9}
\end{equation}

For the FBR of DP the lower limit of the $d\Gamma_B$ of Eq.~(\ref{eq:eC1})
is $\eta_{\min}=2 l p_2\left( y_0-1\right)$, and then $I_0$ is no more
infrared-divergent. We can calculate it from Eq.~(\ref{eq:eC4}) by
changing $\lambda^2$ by $\eta_{\min}$. The second part of $d\Gamma_B$ in
Eq.~(\ref{eq:eC3}) is infrared-convergent. It is more convenient to
transform it to the form of Eq.~(38) of Ref.~\cite{Tun}. We can accomplish
this by performing the $\delta$-integration and by changing the $\eta $
variable to the $y$ variable using $\eta =2p_2 l (y_{0}-y)$. We do not
reproduce the result here because it is very long. It is given in our
Eq.~(\ref{eq:e27}) for $d\Gamma_{B}^\prime$.

\newpage

\begin{table}[h]
\caption{Values of the form factors used in our numerical
calculations. For the first three decays we take the form factor ratios of
Ref.~[6] while for the other decays we use Ref.~[13]. For $\Lambda_c^+
\rightarrow \Lambda$ we take the values of Ref.~[10]. For convenience we
include in the last column the uncorrected $\alpha_\ell^0$ of Eq.~(77)
corresponding to this choice of form factors. }
\begin{tabular}{lddddd}
Decay & $f_{1}$ & $f_{2}$ & $g_{1}$ & $\alpha_{0}^{\ell}$ \\
\tableline
$n\rightarrow P$ & 1.000 & 1.970 & 1.261 & $-$0.0850 \\ 
$\Lambda \rightarrow P$ & 1.236 & 1.199 & 0.890 & 0.0200 \\ 
$\Sigma^-\rightarrow n$ & 1.000 & $-$0.970 & $-$0.340 & $-$0.6319 \\ 
$\Sigma^-\rightarrow \Lambda$ & 0.000 & 1.172 & 0.601 & $-$0.7030 \\ 
$\Sigma^+\rightarrow \Lambda$ & 0.000 & 1.172 & 0.601 & $-$0.6474 \\ 
$\Xi^-\rightarrow \Lambda$ & 1.225 & $-$0.074 & 0.354 & 0.2579 \\ 
$\Xi^-\rightarrow \Sigma^0$ & 0.707 & 1.310 & 0.899 & $-$0.1989 \\ 
$\Xi^0\rightarrow \Sigma^+$ & 1.000 & 1.853 & 1.267 & $-$0.1913 \\ 
$\Lambda_c^+\rightarrow \Lambda$ & 0.350 & 0.090 & 0.610 & 
$-$0.9513 \\
\end{tabular}
\end{table}

\begin{table}[h]
\caption{ Percentage radiative corrections [that is, Eq.~(83) multiplied
by 100] of the spin-asymmetry coefficient of the charged lepton in hyperon
semileptonic decays. The prediction in the fourth column for $
\Lambda_c^+\rightarrow \Lambda$ uses the approach of Ref.~[13], but it was
not actually given there. }
\begin{tabular}{ldddd} 
Decay & $\delta \alpha^{\rm R}=\alpha_{\ell}^{\rm R}-\alpha_{0}^{\ell}$ &
$\delta \alpha^{\rm T}=\alpha_\ell^{\rm T}-\alpha_0^\ell$ & $\delta
\alpha$ Ref.~[13] & $\delta \alpha$ Ref.~[6] \\ 
\tableline 
$n\rightarrow P$ & 0.0119 & 0.0095 & 0.0101 & \\ 
$\Lambda \rightarrow P$ & 0.0813 & 0.0014 & $-$0.0023 & $-$0.0 \\
$\Sigma^-\rightarrow n$ & 0.0832 & 0.0815 & 0.0758 & 0.1 \\
$\Sigma^-\rightarrow \Lambda$ & 0.1432 & 0.0836 & 0.0770 & \\ 
$\Sigma^+\rightarrow \Lambda$ & 0.1287 & 0.0755 & 0.0911 & \\ 
$\Xi^-\rightarrow \Lambda$ & 0.1024 & $-$0.0246 & $-$0.0310 & \\ 
$\Xi^-\rightarrow \Sigma^0$ & 0.3327 & 0.0371 & 0.0212 & \\ 
$\Xi^0\rightarrow \Sigma^+$ & 0.3312 & 0.0350 & 0.0208 & \\ 
$\Lambda_c^+\rightarrow \Lambda$ & 0.0757 & 0.1294 & 0.1098 & \\ 
\end{tabular}
\end{table}

\end{document}